\documentclass[%
 reprint,
 amsmath,amssymb,
aps,
pra,
]{revtex4-2}

\usepackage{graphicx}% Include figure files
\usepackage{dcolumn}% Align table columns on decimal point
\usepackage{bm}% bold math
\usepackage{hyperref}% add hypertext capabilities
\usepackage{xcolor}
\usepackage{soul}

\begin{document}

\preprint{APS/123-QED}

\title{Linear and nonlinear propagation of cylindrical vector beam through a non-degenerate four level atomic system}

\author{Partha Das$^{1,}$}
\email{partha.2015@iitg.ac.in}
\author{Tarak Nath Dey$^{1,}$}
\email{tarak.dey@iitg.ac.in}
\address{$^1$Department of Physics, Indian Institute of Technology Guwahati, Guwahati 781039, Assam, India}


\begin{abstract}
We investigate the phase-induced susceptibilities for both components of the  probe vector beam (PVB) within an atomic system. The atoms are prepared in a non-degenerate four-level configuration. The transitions are coupled by a $\pi$ polarized control field and two orthogonally polarized components of a PVB. We show that the linear susceptibility of the medium depends on the phase shift between the control field and PVB, characterizing loss or gain in the system. Additionally, the phase shift causes polarization rotation in the vector beams (VBs) as they propagate. We further study the effect of nonlinearity on the VB propagation through the medium for a couple of Rayleigh lengths. The self-focusing and defocusing phenomena are observed for radial, azimuthal, and spiral VBs. The special chain-like self-focusing and defocusing leads to the formation of consecutive smaller spot sizes with moderate gain. Therefore, the mechanism of control of susceptibility and self-focusing may hold promise for applications such as transitioning from an absorber to an amplifier, high-resolution microscopy, and optical trap systems.
\end{abstract}

\maketitle

%\tableofcontents

\section{\label{sec:level1}INTRODUCTION}
Increasing interest in tailoring the optical characteristics of a medium stems from its diverse applications in the field of optical physics \cite{intro1}. Prior investigations predominantly centred on examining spatially uniform states of polarization (SOP) to explore quantum coherence, a crucial factor in determining the optical attributes of systems \cite{intro2,intro3,intro4,intro5,intro6}. Quantum coherence is a fundamental requirement for manifesting quantum interference \cite{intro7}. The ideal test bed for realizing quantum interference is a three-level system. Four geometrical level configurations, such as $\Xi$, $V$, $\Lambda$, and $\Delta$, can be formed out of three levels. Apart from these level systems, the double two-level atomic system has been increasingly capturing attention due to the multitude of quantum interference and the corresponding intriguing phenomena within this system. Notable phenomena observed in the double two-level atomic system include spontaneously generated interference in resonance fluorescence \cite{intro8}, efficient modulation of the medium gain for the probe pulse due to interference between the absorption and the stimulated emission paths \cite{intro9}, slowing light through Zeeman coherence oscillations \cite{intro10}, and recovery of interference in resonance fluorescence by relative phase control \cite{intro11}. The coherent control of the effective susceptibility of a duplicated two-level system has been reported in \cite{intro12}. The control can be achieved for a linearly polarized weak field due to the application of a significantly stronger orthogonally polarized field. Furthermore, optical bistability and multistability \cite{intro13}, three-dimensional atom localization \cite{intro14}, and anisotropic nonlinear response through quantum interference \cite{intro15} involves closed-loop double two-level atomic systems.

However, due to their spatially non-uniform polarisation distribution, there is currently a burgeoning fascination with VBs \cite{intro16,intro17,intro18}. A VB or fully structured light (FSL) beam can be generated through the vector superposition of two orthogonally polarized, orbital angular momentum (OAM) carrying Laguerre-Gaussian (LG) modes\cite{intro19}. Two LG modes, possessing equal and opposite OAM can generate a cylindrical vector (CV) beam
that carries a net-zero OAM \cite{intro16}. The polarization distribution of
CV beam exhibits  radial, azimuthal, and spiral with
respect to the axial symmetry of the beam. The polarization distribution, characterized by both radial and azimuthal variations, gives rise to the lemon, star, and web polarization distributions when the two Laguerre-Gaussian (LG) modes possess a non-zero OAM, alias full Poincaré (FP) beams \cite{intro17}.

The recognition of VBs is derived from their adaptability across various domains, where it is useful to retain a desired intensity and polarization distribution\cite{intro20}. CV beams have attracted significant recent attention largely because of their unique properties under high-numerical-aperture (NA) focusing. A radially polarized beam which is a subset of CV beam can generate a strong longitudinal field component at the focal point of a high numerical aperture lens, resulting a tighter spot size \cite{intro21}. This phenomena has been experimentally observed by several groups \cite{intro22,intro23,intro24}. This novel tight focusing property of CV beams has diverse applications in stimulated emission depletion (STED) microscopy, confocal microscopy\cite{intro25,intro26,intro27,intro28}, optical tweezers and manipulation\cite{intro29}. The control of spatial intensity and polarization distribution holds significant importance in various applications such as optical trapping and manipulation\cite{intro30,intro31}, as well as atomic state preparation and detection\cite{intro30,intro32,intro33}. Furthermore, within optical communication, the lateral polarization distribution of vector modes can be leveraged to enhance the data transmission rate of free space optical communication\cite{intro34}. The impact of beam spreading caused by diffraction during linear propagation can be dampened and, in certain instances, precisely balanced by employing a self-focusing (Kerr) nonlinear medium. However, the OAM beams tend to break up into multiple soliton peaks (twice the OAM) during nonlinear propagation\cite{intro35,intro36}. This fragmentation can be suppressed by utilizing VBs instead of scalar beams\cite{intro37,intro38,intro39}.  More recently, based on vector diffraction theory,  focusing characteristics
and pattern changes of axisymmetric Bessel–Gaussian, and  Laguerre–Gaussian (LG) beams have been studied \cite{intro40,intro41}, offering potential applications in the construction of optical traps and optical trap chains, thus holding significant relevance in the field of optical manipulation.

In this work, we have presented phase-induced susceptibilities for both the components of CV beams inside an atomic vapor. The medium is composed of a non-degenerate four-level system which is excited by a strong control field and the two orthogonal polarization components
of a PVB. This non-degenerate system is more realistic as the Zeeman splitting of magnetic sublevels lifts the degeneracy of a duplicated two-level system. After propagating inside the medium, the output probe beam experiences absorption or gain depending on the phase shift between the weak probe and the strong control field. Hashmi {\it et al.}, have achieved the control of susceptibility through wave mixing\cite{intro12}.  The variation in the refractive index of the two components of the CV beam induces polarization rotation. The external magnetic field can control this polarization rotation \cite{intro42}. Furthermore, when the intensity of the probe beam and control field are of comparable magnitude, self-focusing and defocusing in the medium can be achieved by appropriately selecting the detunings of the probe beam and control field transitions. The longitudinal profile of radial, azimuthal, and spiral CV beams show quasiperiodic nature along propagation. In several systems, self-focusing has been observed with scalar beams, including in alkali-metal vapor cell\cite{intro43}, condensate\cite{intro44}, the air on the ground\cite{intro45}, and waveguide arrays\cite{intro46}. However, in our work, self-focusing with CV beams controls the polarization, which is deficient in scalar beams.

The paper is organized as follows. Section I provides a succinct introduction to VBs, their applications, and our research findings. The theoretical formalism adopted in this study is presented in Section II. Section III presents the results of our work, accompanied by comprehensive explanations. Lastly, Section IV includes the paper's conclusion.

\section{THEORETICAL FORMULATION}
\subsection{Level system}
In this paper, we present a scheme based on non-electromagnetically induced gain or absorption arising from manipulating phases of the two components of a vector beam. The system under consideration is a non-degenerate four-level closed-loop system where the relative phase shift between various applied fields can effectively modulate the response for the probe fields. Control fields allow various quantum interference of the electronic excitation pathways that effectively compensate or overcome the linear absorptive spatial response by the non-linear response of the medium. Previous studies on the double two-level system have garnered significant attention due to its wide range of quantum interference phenomena through the application of scalar fields \cite{intro8,intro9,intro10,intro11,intro12,intro13,intro14,intro15}. Motivated by the various aspects of the closed-loop system, we have considered a non-degenerate four-level atomic system driven by two orthogonally
%%%%%%%%%%%%%%%%%%%%%%%%%%%%%%%%%%%%%%%%%%%
%				Figure 1
%%%%%%%%%%%%%%%%%%%%%%%%%%%%%%%%%%%%%%%%%%%
\begin{figure}[ht]
	\centering
	\includegraphics[width=0.7\linewidth]{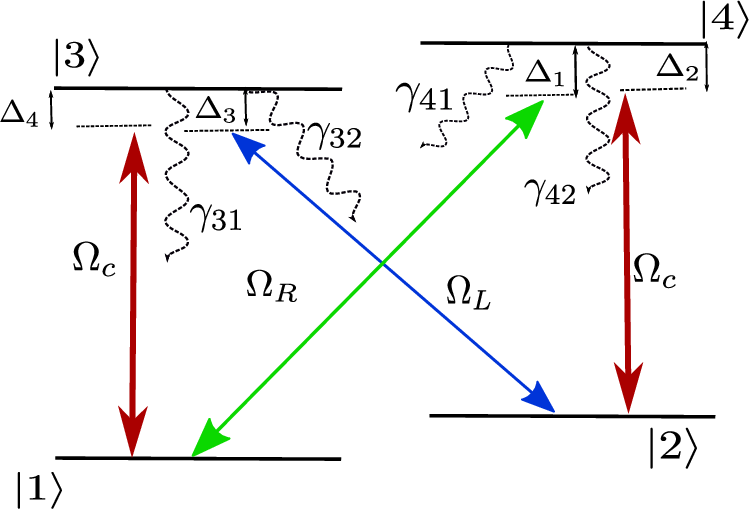}
	\caption{Schematic diagram of a non-degenerate four level atomic system. The right circularly polarized component, $E_R$, and the left circularly polarized component, $E_L$, of a weak probe VB drives the transitions, $|1\rangle\leftrightarrow|4\rangle$ and $|2\rangle\leftrightarrow|3\rangle$ respectively. The transition $|1\rangle\leftrightarrow|3\rangle$ and $|2\rangle\leftrightarrow|4\rangle$ are coupled by a strong control field $E_c$. The spontaneous emission decay rate from $|3\rangle$ and $|4\rangle$ states are given by $\gamma_{3j}$ and $\gamma_{4j}$ ($j\in 1, 2$). The detunings of the transitions are denoted by $\Delta_i$ ($i\in 1,2,3,4$).}
	\label{fig:1}
\end{figure}
%%%%%%%%%%%%%%%%%%%%%%%%%%%%%%%%%%%%%%%%%%%
%%%%%%%%%%%%%%%%%%%%%%%%%%%%%%%%%%%%%%%%%%% 
polarized components of a PVB and a $\pi$ polarized control field as in Fig. \ref{fig:1}. This configuration can be realized in $^6$Li $D_1$($2^2S_{1/2}\rightarrow2^2P_{1/2}$) transition hyperfine structure as: $|1\rangle = |2 ^2S_{1/2}, F = 1/2, m_F = -1/2 \rangle, |2\rangle = |2 ^2S_{1/2}, F = 1/2, m_F = 1/2 \rangle, |3\rangle = |2 ^2P_{1/2}, F = 1/2, m_F = -1/2 \rangle$, and $|4\rangle = |2 ^2P_{1/2}, F = 1/2, m_F = 1/2 \rangle$. The transitions $|1\rangle\leftrightarrow|3\rangle$, and $|2\rangle\leftrightarrow|4\rangle$ are coupled by a $\pi$-polarized control field $\Vec{E}_c$, which is defined as,
%%%%%%%%%%%%%%%%%%%%%%%%%%%%%%%%%%%%%%%%%%%
\begin{align}
&\Vec{E}_c(r,t) = \hat{e}_{\pi} \mathcal{E}_c(r) e^{-i(\omega_ct - k_cz)} + c.c.,
\end{align} 
%%%%%%%%%%%%%%%%%%%%%%%%%%%%%%%%%%%%%%%%%%%
where $\hat{e}_{\pi}$, $\mathcal{E}_c(r)$, $\omega_c$, $k_c$ are the polarization vector, spatial envelop, central frequency and wavevector respectively. We can decompose a linear polarized vector probe field $\Vec{E}_p$, into two orthogonally polarized basis states $\hat{\sigma}_i(i\in R,L)$ as,
%%%%%%%%%%%%%%%%%%%%%%%%%%%%%%%%%%%%%%%%%%%
\begin{subequations}
	\begin{align}
		\Vec{E}_p(r,t) &= \hat{e}_{x} \mathcal{E}_p(r) e^{-i(\omega_pt - k_pz + \beta)} + c.c.\\
		&=\sum_{i=R,L}\hat{\sigma_i} \mathcal{E}_i(r)e^{-i(\omega_pt - k_pz + \beta)} + c.c.,
	\end{align} 
\end{subequations}
%%%%%%%%%%%%%%%%%%%%%%%%%%%%%%%%%%%%%%%%%%%
where $\hat{\sigma}_{R(L)}$ are the right (left) circular polarization unit vector, the right(left) circular polarized component $\mathcal{E}_{R(L)}$ components couples with the $|1\rangle\leftrightarrow|4\rangle$ $(|2\rangle\leftrightarrow|3\rangle)$. The phase shift among the probe and the control fields is $\beta$.

The time-dependent Hamiltonian describing the interaction of the model system as shown in Fig. \ref{fig:1}, can be written under dipole approximation as
\begin{subequations}
	\begin{align}
		\textbf{\textit{H}} =& \textbf{\textit{H}}_{0}+ \textit{\textbf{H}}_{I}, \\
		\textbf{\textit{H}}_{0} =&  \hbar (\omega_{21} |2 \rangle \langle 2 | + \omega_{31} |3 \rangle \langle 3 | + \omega_{41} |4 \rangle \langle 4 |) ,\\
		\textit{\textbf{H}}_{I} =& - \hat{d}.\vec{E}\nonumber\\
		=& -[\vec{d}_{41}.(\hat{e}_R \mathcal{E}_R e^{-i\omega_p t - i\beta} + c.c.)|4\rangle \langle 1|\nonumber\\&+\vec{d}_{32}.(\hat{e}_L \mathcal{E}_L e^{-i\omega_p t - i\beta} + c.c.)|3\rangle \langle 2|\nonumber\\&+\vec{d}_{31}.(\hat{e}_{\pi} \mathcal{E}_c e^{-i\omega_c t} + c.c.)|3\rangle \langle 1|\nonumber\\&+\vec{d}_{42}.(\hat{e}_{\pi} \mathcal{E}_c e^{-i\omega_c t} + c.c.)|4\rangle \langle 2|] + \text{H.c.,}
    \end{align}
\end{subequations}
where $\omega_{j1}$($j = 2, 3, 4$) correspond to the frequency separation between the state $|j\rangle$ and the ground state $|1\rangle$ and $\vec{d}_{ik}$ ($i = 3, 4$; $k = 1, 2$) are the matrix elements of the induced dipole moments for the transitions $|i\rangle\leftrightarrow |k\rangle$. In order to eliminate the explicit time dependence in the Hamiltonian, we perform the following unitary transformation
\begin{equation}
	U = \text{exp}[-i\omega_c t |3\rangle\langle 3| - i\omega_p t |4\rangle\langle 4| - i(\omega_p-\omega_c)t|2\rangle\langle 2|].
\end{equation}
Now the effective Hamiltonian in the interaction picture is given as, $\mathcal{H} = \hat{U}^\dagger H \hat{U} - i\hbar\hat{U}^\dagger\partial_t \hat{U}$. We impose the condition $\omega_p - \omega_c = 0$, so that the time dependence is completely eliminated from the effective Hamiltonian. Under the rotating wave approximation (RWA) it gives
\begin{align}
	\mathcal{H} =& -\hbar[(\Delta_1-\Delta_2)|2\rangle\langle 2|+\Delta_4|3\rangle\langle 3|+\Delta_1|4\rangle\langle 4|]\nonumber\\&-\hbar[\Omega_c|3\rangle\langle 1| -\Omega_L\text{e}^{-i\beta}|3\rangle\langle 2|+\Omega_R\text{e}^{-i\beta}|4\rangle\langle 1|\nonumber\\&-\Omega_c|4\rangle\langle 2|] + \text{H.c.}
\end{align}
The single photon detunings of the probe and control field for their respective transitions are defined as
\begin{align}
	&\Delta_1 = \omega_p - \omega_{41},\hspace{0.1cm} \Delta_2 = \omega_c - \omega_{42},\hspace{0.1cm} \Delta_3 = \omega_p - \omega_{32},\nonumber\\&\Delta_4 = \omega_c - \omega_{31},
\end{align}
and the Rabi frequencies of probe field components and control field is written as
\begin{align}
    &\Omega_R = \dfrac{\vec{d}_{41}. \hat{e}_R}{\hbar}\mathcal{E}_R, \hspace{0.2cm}  \Omega_L = \dfrac{\vec{d}_{32}. \hat{e}_L}{\hbar}\mathcal{E}_L, \hspace{0.2cm}  \Omega_c = \dfrac{\vec{d}_{31}. \hat{e}_\pi}{\hbar}\mathcal{E}_c .
\end{align}
Note that $\vec{d}_{32} = -\vec{d}_{41}$, and $\vec{d}_{42} = -\vec{d}_{31}$ according to the Clebsch-Gordan coefficient for the $F = 1/2 \leftrightarrow F =1/2$ level scheme. To find the population dynamics and the atomic coherence we use the following Liouville equation 
\begin{equation}
		\dfrac{\partial \rho}{\partial t} = -\dfrac{i}{\hbar}[\mathcal{H},\rho] + \mathcal{L} \rho.
\end{equation}
The second term on the right hand side of the above equation represents all the radiative and nonradiative decay processes which can be determined by
\begin{equation}
	\mathcal{L}\rho = -\sum_{i=3}^{4}\sum_{j=1}^{2}\dfrac{\gamma_{ij}}{2}(|i\rangle\langle i|\rho - 2|j\rangle\langle j|\rho_{ii} + \rho|i\rangle\langle i|),
\end{equation}
where $\gamma_{ij}$ are the radiative decay rates from the excited state $|i\rangle$ to $|j\rangle$. 

Now, the dynamics of the model system can be obtained from the following density matrix equations
\begin{subequations}\label{eq.10}
	\begin{align}
		\dot{\rho}_{11} =& -i(\Omega_c\rho_{13} + \Omega_R\text{e}^{-i\beta}\rho_{14}-\Omega_c^*\rho_{31}-\Omega_R^*\text{e}^{i\beta}\rho_{41})\nonumber\\&+\gamma_{31}\rho_{33}+\gamma_{41}\rho_{44},\\
		\dot{\rho}_{22} =& -i(\Omega_L\text{e}^{-i\beta}\rho_{23}-\Omega_c\rho_{24}-\Omega_L^*\text{e}^{i\beta}\rho_{32}+\Omega_c^*\rho_{42})\nonumber\\&+\gamma_{32}\rho_{33}+\gamma_{42}\rho_{44},\\
		\dot{\rho}_{33} =& i(\Omega_c\rho_{13}+\Omega_L\text{e}^{-i\beta}\rho_{23}-\Omega_c^*\rho_{31}-\Omega_L^*\text{e}^{i\beta}\rho_{32})\nonumber\\&-(\gamma_{31}+\gamma_{32})\rho_{33},\\
		\dot{\rho}_{31} =& i(\Omega_L\text{e}^{-i\beta}\rho_{21}+\Delta_{31}\rho_{31}+\Omega_c(\rho_{11}-\rho_{33})\nonumber\\&-\Omega_R\text{e}^{-i\beta}\rho_{34}),\\
		\dot{\rho}_{32} =& -i(\Delta_{32}\rho_{32}+\Omega_L\text{e}^{-i\beta}(\rho_{33}-\rho_{22})-\Omega_c(\rho_{12}+\rho_{34})),\\
		\dot{\rho}_{41} =& i(\Delta_{41}\rho_{41}-\Omega_c(\rho_{21}+\rho_{43})+\Omega_R\text{e}^{-i\beta}(\rho_{11}-\rho_{44})),\\
		\dot{\rho}_{42} =& i(\Omega_R\text{e}^{-i\beta}\rho_{12}-\Omega_L\text{e}^{-i\beta}\rho_{43}+\Delta_{42}\rho_{42}+\Omega_c(\rho_{44}\nonumber\\&-\rho_{22})),\\
	    \dot{\rho}_{34} =& i(\Omega_c\rho_{14}+\Omega_L\text{e}^{-i\beta}\rho_{24}-\Omega_R^*\text{e}^{i\beta}\rho_{31}+\Omega_c^*\rho_{32}\nonumber\\&+\Delta_{34}\rho_{34}),\\
		\dot{\rho}_{12} =& -i(\Delta_{12}\rho_{12}+\Omega_L\text{e}^{-i\beta}\rho_{13}-\Omega_c\rho_{14}-\Omega_c^*\rho_{32}\nonumber\\&-\Omega_R^*\text{e}^{i\beta}\rho_{42}),
	   \end{align}
\end{subequations}
where the overdots signify time derivatives and star ($*$) denotes the complex conjugates. The remaining density matrix equations are derived from the population conservation law $\sum_{i=1}^{4}\rho_{ii}=1$ and their complex conjugate expressions $\dot{\rho}_{ij} = \dot{\rho}_{ji}^*$. In Eq.(\ref{eq.10}), $\Delta_{31} = \Delta_4+i\Gamma_{31}$, $\Delta_{32} = (\Delta_1-\Delta_2-\Delta_4)-i\Gamma_{32}$, $\Delta_{41} = \Delta_1+i\Gamma_{41}$, $\Delta_{42} = \Delta_2+i\Gamma_{42}$, $\Delta_{34} = (\Delta_4-\Delta_1)+i\Gamma_{34}$, $\Delta_{12} = (\Delta_1-\Delta_2)-i\Gamma_{12}$ and $\Gamma_{ij}$ are the decoherence rates of the system.
\subsection{Linear Response of the medium}
In this section, we present an analytical expression for the atomic coherence to validate the linear response of the PVB within the medium. The PVB is to be sufficiently weak ($|\Omega_{L,R}|<<|\Omega_{c}|$) to consider it as a peturbation to a system of linear order under the steady state condition. The perturbative expansion of the density matrix upto the first order in the probe beam can be expressed as \cite{lr1}
%####################################################
\begin{align}\label{eq.11}
	\rho_{ij} = \rho_{ij}^{(0)} + \Omega_L \rho_{ij}^{(1L)} + \Omega_L^* \rho_{ij}^{(1L^*)} + \Omega_R \rho_{ij}^{(1R)} + \Omega_R^* \rho_{ij}^{(1R^*)} ,
\end{align}
%####################################################
where $\rho_{ij}^{(0)}$ describes the solution in absence of the PVB, and $\rho_{ij}^{(1k)}$ ($k\in L,L^*,R,R^*$) represent the linear order solutions in presence of the PVB. To find the coherence, we substitute Eq. (\ref{eq.11}) in Eqs. (\ref{eq.10}) and equate the coefficients of $\Omega_L$, $\Omega_L^*$, $\Omega_R$, $\Omega_R^*$. Thus we obtain four sets of fifteen coupled linear equations. We then solve these algebraic equations to obtain the coherence which is given as
%####################################################
\begin{subequations}\label{eq.12}
		\begin{align}
			&\rho_{32}^{1L} = -\frac{i e^{-i \beta } (A+B)}{C D},\\
			&\rho_{32}^{1R^*} = -\frac{ie^{i \beta }E}{C D},\\
			&\rho_{41}^{1R} = -\frac{i e^{-i \beta } (F+G)}{DC^*},\\
			&\rho_{41}^{1L^*} = -\frac{ie^{i \beta }H}{DC^*}.
		\end{align}
\end{subequations}
%####################################################
In Eqs. (\ref{eq.12}) the detailed expressions for $A-H$ are provided in the Appendix. It is pertinent to note that in the original frame of reference, the off-diagonal density matrix element $\rho_{32}$ and  $\rho_{41}$ must be multiplied by $e^{-i\omega_p t}$. The steady-state values of $\rho_{32}$ and $\rho_{41}$ determine the linear susceptibility $\chi_{32}$ and $\chi_{41}$ of the medium at the frequency $\omega_p$, respectively. We consider both the probe components are at two photon resonance, with $\Delta_1 = \Delta_2 = \Delta_R$ and $\Delta_3 = \Delta_4 = \Delta_L $. The decay rates, $\gamma_{31} = \gamma_{42} = \gamma/3$, $\gamma_{41} = \gamma_{32} = 2\gamma/3$, and coherence decay between the states $|2\rangle$ and $|1\rangle$ being negligible, the medium's susceptibility can be expressed as follows:
%####################################################
\begin{widetext}
\begin{subequations}\label{eq.13}
	\begin{align}
		&\chi_{32}(\omega_p) = \frac{\mathcal{N}|d_{32}|^2}{\hbar}\dfrac{ie^{i\beta}}{\Gamma_2}\biggl[\Gamma_1 \biggl(\dfrac{\Omega_c}{\Omega_c^*}\biggl)\biggl(\dfrac{\Omega_R^*}{\Omega_L}\biggl) + \Gamma_3 e^{-2i\beta}\biggl],\\
		&\chi_{41}(\omega_p) = \frac{\mathcal{N}|d_{41}|^2}{\hbar}\dfrac{ie^{i\beta}}{\Gamma_5}\biggl[\Gamma_4 \biggl(\dfrac{\Omega_c}{\Omega_c^*}\biggl)\biggl(\dfrac{\Omega_L^*}{\Omega_R}\biggl) + \Gamma_6 e^{-2i\beta}\biggl],
	\end{align}
\end{subequations}
where,
\begin{subequations}\label{eq.14}
\begin{align}
	&\Gamma_1 = \Gamma_R^*-\Gamma_L\Gamma_L^*;\hspace{.5cm}\Gamma_2 =- (\Gamma_L+\Gamma_R^*)(4|\Omega_c|^2+\Gamma_L\Gamma_L^*+\Gamma_R\Gamma_R^*);\hspace{.5cm}\Gamma_3=i\Gamma_R^*(\Delta_R - \Delta_L);\\
	&\Gamma_4 = \Gamma_L^*-\Gamma_R\Gamma_R^*;\hspace{.5cm}\Gamma_5 =- (\Gamma_R+\Gamma_L^*)(4|\Omega_c|^2+\Gamma_L\Gamma_L^*+\Gamma_R\Gamma_R^*);\hspace{.5cm}\Gamma_6=i\Gamma_L^*(\Delta_L - \Delta_R);\\
	&\Gamma_R = (i\Delta_R - 1/2);\hspace{.3cm}\Gamma_L = (i\Delta_L - 1/2).
\end{align}
\end{subequations} 
\end{widetext}
%####################################################
Here $\mathcal{N}$ is the atomic density of the medium. The expression for
%%%%%%%%%%%%%%%%%%%%%%%%%%%%%%%%%%%%%%%%%%%
%				Figure 2
%%%%%%%%%%%%%%%%%%%%%%%%%%%%%%%%%%%%%%%%%%%
\begin{figure}[ht]
	\centering
	\includegraphics[width=\linewidth]{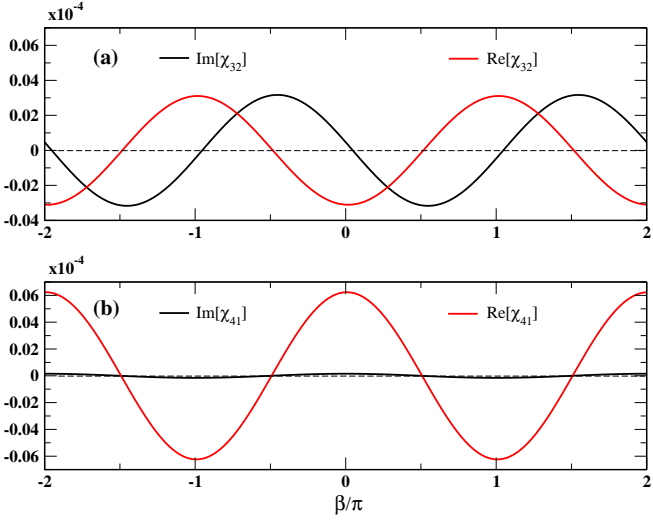}
	\caption{(a) Real and imaginary part of $\chi_{32}$ vs. phase shift $\beta$. (b) Real and imaginary part of $\chi_{41}$ vs. phase shift $\beta$.  Parameter used: $|\Omega_{R,L}|=0.02\gamma,\Omega_{c}=2\gamma$. $\gamma_{31} = \gamma_{42} = \gamma/3$, $\gamma_{41} = \gamma_{32} = 2\gamma/3$. $\Delta_L = 0, \Delta_R =10\gamma$. The density of atoms, $\mathcal{N} = 2\times10^{10}\text{cm}^{-3}$.}
	\label{fig.2}
\end{figure} 
%%%%%%%%%%%%%%%%%%%%%%%%%%%%%%%%%%%%%%%%%%%
the susceptibilities of the transitions  $|2\rangle \leftrightarrow |3\rangle$, and $|1\rangle \leftrightarrow |4\rangle$, as described herein, provide valuable insights into the physics of phase-dependent interference. Under two-photon resonance conditions, the real and imaginary parts of susceptibility for both the VB components are exhibited in Figs. \ref{fig.2}(a) and \ref{fig.2}(b). The positive and negative value of Im[$\chi_{32}$], and Im[$\chi_{41}$] signifies absorption and gain of the medium, respectively. This oscillatory nature as a function of the phase shift $\beta$ corresponds to different quantum pathways in the closed-loop system. On the other hand, the difference in the Re[$\chi_{32}$] and Re[$\chi_{41}$] has a direct impact on the polarization rotation of the CV beam. In the latter part of this paper, the phase shift $\beta = 0$ and $\beta = \pi/2$  are adopted. The phase shift $\beta$ significantly impacts the redistribution of population and consequential changes in atomic coherence
within our closed-loop system. As such, it serves as a control mechanism, resulting in either constructive or destructive interference among various excitation pathways. This feature effectively acts as a control knob for the system, enabling precise regulation of absorption and dispersion. To investigate the medium's nonlinear response, we use the Gaussian elimination method to find the relevant coherences for a probe field in all orders of probe and control fields at higher intensity limits.
%%%%%%%%%%%%%%%%%%%%%%%%%%%%%%%%%%%%%%%%%%%
\subsection{Stokes parameter}
The four Stokes polarization parameters are required to describe any polarization state of light. The Stokes parameters are precisely the quantities that are experimentally measured. To understand the SOP in the transverse plane of the VB, we decompose VB into its two orthogonally polarized LG modes. We use a circular polarization basis of the VB that can be emerging as follows:
%%%%%%%%%%%%%%%%%%%%%%%%%%%%%%%%%%%%%%%%%%%
\begin{equation}\label{eq.15}
 	\vec{E}(r,\phi,z) = \mathcal{E}_L(r,\phi,z)\hat{e}_L + \mathcal{E}_R(r,\phi,z)\hat{e}_R,
\end{equation}
where,
\begin{equation}\label{eq.16}
	\mathcal{E}_L(r,\phi,z) = \text{cos}(\alpha)LG_0^{l_L}, \hspace{0.2cm} \mathcal{E}_R(r,\phi,z) = \text{sin}(\alpha)e^{i\theta}LG_0^{l_R}.
\end{equation}
%%%%%%%%%%%%%%%%%%%%%%%%%%%%%%%%%%%%%%%%%%%
$\mathcal{E}_L $, and  $\mathcal{E}_R$ are the left and right circularly polarized component of the VB. The variables $\alpha$, and $\theta$ are the relative amplitude and phase of the two modes, respectively. The spatial modes, $LG_0^{l_i}(i = L, R)$ are the Laguerre Gaussian polynomial, with the radial index zero is given by 
%%%%%%%%%%%%%%%%%%%%%%%%%%%%%%%%%%%%%%%%%%%
\begin{align}\label{eq.17}
	LG_0^{l_i}(r, \phi, z) =& \mathcal{E}_0 \sqrt{\dfrac{2}{\pi|l_i|!}} \left(\dfrac{r\sqrt{2}}{w(z)}\right)^{|l_i|}e^{\scalebox{0.8}{$-\dfrac{r^2}{w(z)^2}$}}e^{il_i\phi+ik_in_iz}\nonumber \\ &\times  \text{exp}\left(\dfrac{ik_in_ir^2z}{2(z^2+n_i^2z_R^2)}\right)e^{-i(|l_i| + 1)\eta(z)}.
\end{align}
%%%%%%%%%%%%%%%%%%%%%%%%%%%%%%%%%%%%%%%%%%%
In Eq. (\ref{eq.17}) the beam radius at a propagation length $z$ is denoted by $w(z) = w_0\sqrt{1+z^2/n_i^2z_R^2}$, where $w_0$ is the beam waist at $z = 0$, and $n_i$ is the refractive index. The free space Rayleigh length is $z_R = k_iw_0^2/2$, with $k_i$ being the free space wave number. The OAM index is represented by $l_i$, while the Gouy phase can be expressed as  $(|l_i| + 1)\eta(z)$, where $\eta(z) = \text{tan}^{-1}(z/n_iz_R)$. The refractive indices, $n_R = 1+2\pi \text{Re}[\chi_{41}]$, and $n_L = 1+2\pi \text{Re}[\chi_{32}]$. In Gaussian units, the susceptibilities can also be expressed as $\chi_{41}=\eta'\rho_{41}/\Omega_{R}$, and $\chi_{32}=\eta'\rho_{32}/\Omega_{L}$. Here the parameter $\eta' = 3\mathcal{N}/2k_i^3$ $(i=L, R)$, which is dimensionless. Now if the two modes of the VB have equal amplitude and equal but opposite OAM the resulting CV beam will have spatially varying linear polarization which can be radial, azimuthal, and sprial depending on the phase difference between the two modes. In case of unequal relative amplitude it causes elliptical polarization distribution. The Stokes parameters in the circular basis is given as
%%%%%%%%%%%%%%%%%%%%%%%%%%%%%%%%%%%%%%%%%%%
\begin{align}\label{eq.18}
		&S_0 = |\mathcal{E}_R|^2 + |\mathcal{E}_L|^2,\hspace{0.2cm} S_1 = 2\text{Re}[\mathcal{E}_R^* \mathcal{E}_L],\nonumber\\ 
		&S_2 = 2\text{Im}[\mathcal{E}_R^* \mathcal{E}_L],\hspace{0.2cm} S_3 = |\mathcal{E}_L|^2 - |\mathcal{E}_R|^2.
\end{align}
%%%%%%%%%%%%%%%%%%%%%%%%%%%%%%%%%%%%%%%%%%%
From  Eq. (\ref{eq.18}), we can calculate the ellipticity, $\zeta$ and the orientation, $\xi$ of polarization at each point in the transverse plane as
%%%%%%%%%%%%%%%%%%%%%%%%%%%%%%%%%%%%%%%%%%%
\begin{align}\label{eq.19}
	&\dfrac{S_1}{S_0} = \text{cos}(2\zeta)\text{cos}(2\xi),\hspace{0.2cm} \dfrac{S_2}{S_0} = \text{cos}(2\zeta)\text{sin}(2\xi)\nonumber\\& \dfrac{S_3}{S_0} = \text{sin}(2\zeta),
\end{align}
%%%%%%%%%%%%%%%%%%%%%%%%%%%%%%%%%%%%%%%%%%%
which give
%%%%%%%%%%%%%%%%%%%%%%%%%%%%%%%%%%%%%%%%%%%
\begin{align}\label{eq.20}
	&\zeta = \dfrac{1}{2} \text{sin}^{-1}\left(\dfrac{S_3}{S_0}\right), \hspace{0.2cm} \xi = \dfrac{1}{2} \text{tan}^{-1}\left(\dfrac{S_2}{S_1}\right).
\end{align} 
%%%%%%%%%%%%%%%%%%%%%%%%%%%%%%%%%%%%%%%%%%%
We substitute Eqs. (\ref{eq.16}), (\ref{eq.17}), and (\ref{eq.18}) in Eq. (\ref{eq.20}) assuming that the free space wave vector $k_R = k_L = k$, gives
%%%%%%%%%%%%%%%%%%%%%%%%%%%%%%%%%%%%%%%%%%%
\begin{align}\label{eq.21}
	\xi(z) =& -\dfrac{1}{2}\biggl[\theta + \phi\Delta(l_{L,R}) + kz\Delta(n_{R,L})+\eta(z)\Delta(|l_{L,R}|)\nonumber\\&+
	\dfrac{kzr^2}{2}\biggl\{\dfrac{n_R}{z^2+n_R^2z_R^2}-\dfrac{n_L}{z^2+n_L^2z_R^2}\biggl\}\biggl],
\end{align}
%%%%%%%%%%%%%%%%%%%%%%%%%%%%%%%%%%%%%%%%%%%
where $\Delta(l_{L,R}) = l_L-l_R$, $\Delta(|l_{L,R}|) = |l_L|-|l_R|$, and $\Delta(n_{R,L}) = n_R-n_L$. The two components of a CV beam have equal and opposite OAM index, which implies $|l_R|=|l_L|$. Therefore, after propagating a distance $z$ through the medium, the polarization of a CV beam at each point on the transverse plane rotates by an amount which is given by
%%%%%%%%%%%%%%%%%%%%%%%%%%%%%%%%%%%%%%%%%%%
\begin{align}\label{eq.22}
	\Delta\xi(z) =& -\dfrac{1}{2}\biggl[
	\dfrac{kzr^2}{2}\biggl\{\dfrac{n_R}{z^2+n_R^2z_R^2}-\dfrac{n_L}{z^2+n_L^2z_R^2}\biggl\}\nonumber\\&+kz\Delta(n_{R,L})\biggl].
\end{align}
%%%%%%%%%%%%%%%%%%%%%%%%%%%%%%%%%%%%%%%%%%%
According to the Eq. (\ref{eq.22}), it is evident that the polarization rotation inside the medium in the case of CV beams is attributed solely to the difference in the refractive index of the two components of the probe beam. Additionally, we can also express the variation of ellipticity for the CV beam as
%%%%%%%%%%%%%%%%%%%%%%%%%%%%%%%%%%%%%%%%%%%
\begin{align}\label{eq.23}
	\Delta\zeta(z) =& \dfrac{1}{2}\biggl[\text{sin}^{-1}\biggl\{
	\dfrac{1 - a \text{tan}^2\alpha}{1 + a \text{tan}^2\alpha}\biggl\}- \text{sin}^{-1}\biggl\{\dfrac{1 -  \text{tan}^2\alpha}{1 +  \text{tan}^2\alpha}\biggl\}\biggl],
\end{align}
%%%%%%%%%%%%%%%%%%%%%%%%%%%%%%%%%%%%%%%%%%%
where,
%%%%%%%%%%%%%%%%%%%%%%%%%%%%%%%%%%%%%%%%%%%
\begin{align}\label{eq.24}
	a=&\text{exp}\biggl[-\dfrac{2r^2}{w_0^2}\left(\dfrac{n_R^2z_R^2}{z^2+n_R^2z_R^2}-\dfrac{n_L^2z_R^2}{z^2+n_L^2z_R^2}\right)\biggl]\nonumber \\ &\times\sqrt{\dfrac{n_R^2(z^2+n_L^2z_R^2)}{n_L^2(z^2+n_R^2z_R^2)}}.
\end{align}
%%%%%%%%%%%%%%%%%%%%%%%%%%%%%%%%%%%%%%%%%%%
\subsection{Propagation equation}
The study of beam propagation equations is crucial to explore the effect of absorption, diffraction, dispersion, and anisotropicity on VB propagation. Under the slowly varying envelope and paraxial wave approximations, the propagation equations for the right and left circularly polarized components of the probe VB can be expressed as
%%%%%%%%%%%%%%%%%%%%%%%%%%%%%%%%%%%%%%%%%%%
\begin{subequations}{\label{eq.25}}
\begin{align}
		&\dfrac{\partial\Omega_R}{\partial z} = \dfrac{i}{2k_R}\nabla^2_{\perp}\Omega_R+\dfrac{2\pi i k_R\mathcal{N}|d_{14}|^2}{\hbar}\rho_{41},\\
		&\dfrac{\partial\Omega_L}{\partial z} = \dfrac{i}{2k_L}\nabla^2_{\perp}\Omega_L+\dfrac{2\pi i k_L\mathcal{N}|d_{23}|^2}{\hbar}\rho_{32}.
\end{align}
\end{subequations}
%%%%%%%%%%%%%%%%%%%%%%%%%%%%%%%%%%%%%%%%%%%
The first term on the right-hand side of the above equation is accountable for diffraction, and the second term represents the dispersion and absorption of the probe VB. Moderate gain or low absorption is required to propagate a VB through the medium so that the gain or absorption of the medium cannot distort the beam's spatial profile. The propagation dynamics of the strong control fields $\vec{E}_c$ are ignored due to their geometrical structure and polarisation characteristics. The split-step Fourier method (SSFM) has been chosen to study the Eqs. (\ref{eq.25}a) and (\ref{eq.25}b)  numerically. In instances with a weak probe and a strong control field, the resulting outcomes align perfectly with the coherences calculated via the perturbation technique, as outlined in Eq. (\ref{eq.11}).
%%%%%%%%%%%%%%%%%%%%%%%%%%%%%%%%%%%%%%%%%%%
%%%%%%%%%%%%%%%%%%%%%%%%%%%%%%%%%%%%%%%%%%%
\section{Numerical Results}
\subsection{Control of linear susceptibility}
In this section, we have explored phase-induced medium susceptibilities for both components of the probe field
%%%%%%%%%%%%%%%%%%%%%%%%%%%%%%%%%%%%%%%%%%%
%				Figure 3
%%%%%%%%%%%%%%%%%%%%%%%%%%%%%%%%%%%%%%%%%%%
\begin{figure}[ht]
	\centering
	\includegraphics[width=\linewidth]{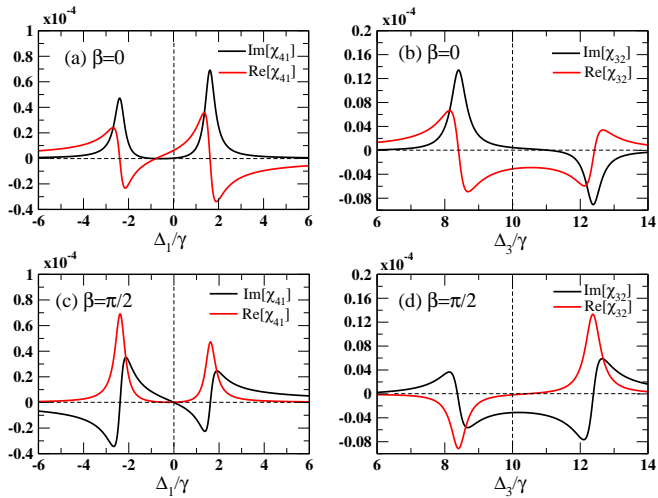}
	\caption{Real and imaginary part of $\chi_{41}$ and $\chi_{32}$ is plotted against the corresponding transition detunings. (a), (b) correspond to $\beta = 0$, plotted against $\Delta_1/\gamma$ and $\Delta_3/\gamma$ respectively. Similarly (c), (d) correspond to $\beta = \pi/2$, plotted against $\Delta_1/\gamma$ and $\Delta_3/\gamma$ respectively. Parameter used: $|\Omega_{R,L}|=0.02\gamma,\Omega_{c}=2\gamma$. $\gamma_{31} = \gamma_{42} = \gamma/3$, $\gamma_{41} = \gamma_{32} = 2\gamma/3$. $\Delta_2 = 0, \Delta_4 =10\gamma$. The density of atoms, $\mathcal{N} = 2\times10^{10}\text{cm}^{-3}$.}
	\label{fig.3}
\end{figure} 
%%%%%%%%%%%%%%%%%%%%%%%%%%%%%%%%%%%%%%%%%%%
that involve two-photon resonance conditions as in Fig. \ref{fig.2}. It is evident from Figs. \ref{fig.3}(a)-(d) how the absorption (or gain), and dispersion are sensitive due to the phase shift between the probe and control field at two different values of $\beta = 0$, and $\beta = \pi/2$.
%%%%%%%%%%%%%%%%%%%%%%%%%%%%%%%%%%%%%%%%%%%
%				Figure 4
%%%%%%%%%%%%%%%%%%%%%%%%%%%%%%%%%%%%%%%%%%%
\begin{figure}[ht]
	\centering
	\includegraphics[width=\linewidth]{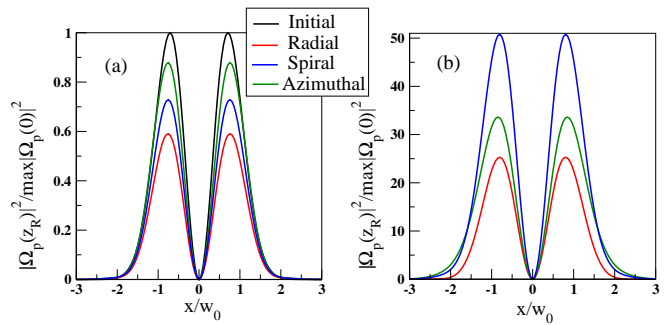}
	\caption{Normalized probe intensity is plotted against the transverse coordinate for different values of $\beta$. (a) For $\beta = 0$ the medium shows absorption, and (b) for $\beta = \pi/2$ it exhibits gain. The paper maintains a consistent beam waist of $w_0 = 60\mu$m, throughout. We choose $\Delta_1 =0$ and the other parameters are same as Fig. \ref{fig.3}.}
	\label{fig.4}
\end{figure}
%$$$$$$$$$$$$$$$$$$$$$$$$$$$$$$$$$$$$$$$$$$$$$$$$$$$$$$$$$$$$$$$$$
The susceptibility at two-photon resonance for both VB components precisely matches our analytical result presented in Fig. \ref{fig.2}. The asymmetric nature at the two-photon resonance condition ($\Delta_1 =0$ or $\Delta_3 =10\gamma$) comes from significant positive detuning $\Delta_4$ of the control field. Depending on the phase shift, the response of the medium on the CV beams is shown in Figs. \ref{fig.4}(a) and \ref{fig.4}(b) at a  
%$$$$$$$$$$$$$$$$$$$$$$$$$$$$$$$$$$$$$$$$$$$$$$$$$$$$$$$$$$$$$$$$$
%%%%%%%%%%%%%%%%%%%%%%%%%%%%%%%%%%%%%%%%%%%
%				Figure 5
%%%%%%%%%%%%%%%%%%%%%%%%%%%%%%%%%%%%%%%%%%%
\begin{figure}[ht]
	\centering
	\includegraphics[width=\linewidth]{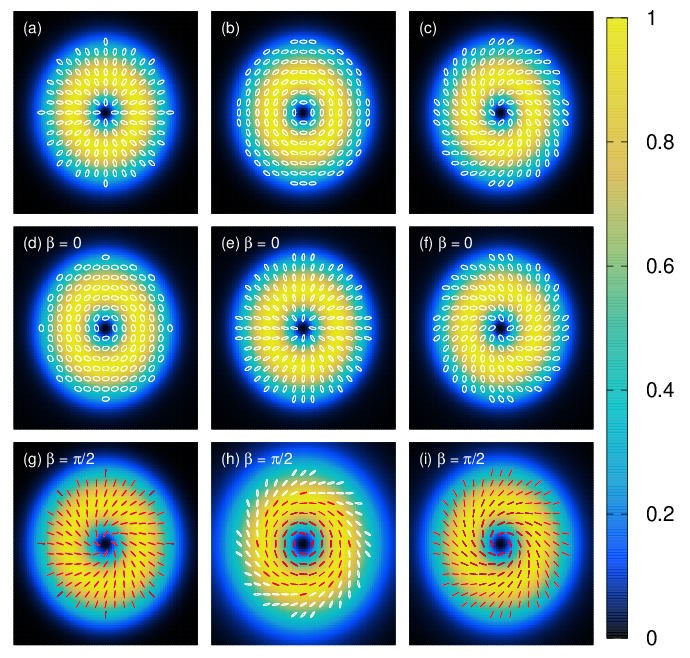}
	\caption{Transverse intensity and polarization distribution of CV beams($l_L = -1, l_R = 1$). (a) Radial($\alpha = \pi/8, \theta = 0$), (b) azimuthal($\alpha = \pi/8, \theta = \pi$), (c) spiral($\alpha = \pi/8, \theta = -\pi/2$) at $z = 0$. (d)-(f) For $\beta = 0$ (radial, azimuthal, and spiral respectively) at $z = z_R$. (g)-(i) For $\beta = \pi/2$ (radial, azimuthal, and spiral respectively) at $z = z_R$. The colors white, and red correspond to left circular, linear polarizations, respectively. The other parameters are same as Fig. \ref{fig.3}.}
	\label{fig.5}
\end{figure}
%%%%%%%%%%%%%%%%%%%%%%%%%%%%%%%%%%%%%%%%%%%
%				Figure 6
%%%%%%%%%%%%%%%%%%%%%%%%%%%%%%%%%%%%%%%%%%%
\begin{figure*}
	\centering
	\includegraphics[width=0.9\linewidth]{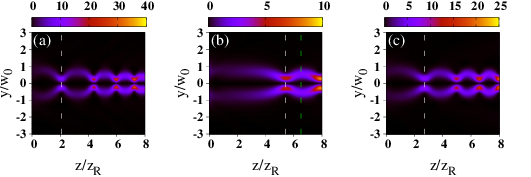}
	\caption{Longitudinal intensity profile of (a) radial, (b) azimuthal, and (c) sprial CV beams. Parameter used: $|\Omega_{R,L}|=3\gamma,\Omega_{c}=4\gamma$. $\Delta_1 = -1.8\gamma, \Delta_2 = 0.15\gamma, \Delta_4 = 13\gamma$, $\beta$ = 0, $\mathcal{N} = 4\times10^{11}\text{cm}^{-3}$. White dashed vertical line denotes the first focal point for each beam, and the green dashed line represents a typical defocused point for azimuthal CV beam.}
	\label{fig.6}
\end{figure*}
%%%%%%%%%%%%%%%%%%%%%%%%%%%%%%%%%%%%%%%%%%% 
%%%%%%%%%%%%%%%%%%%%%%%%%%%%%%%%%%%%%%%%%%% 
distance $z_R$. Note that we work in the paraxial regime throughout the paper, which is valid since our beam waist is much larger than the wavelength. Moreover, the vector superposition of the LG beam corresponds to the solution of the paraxial wave equation in cylindrical coordinates. In the case of $\beta = 0$, we observe absorption for the CV beams such as radial, azimuthal and spiral, which is illustrated in Fig. \ref{fig.4}(a). Certainly, the destructive interference between the quantum pathways causes this phenomenon. However, for $\beta = \pi/2$, the medium shows substantial gain for all three kinds of CV beams, as shown in Fig. \ref{fig.4}(b). This is attributed to the constructive interference phenomena. Therefore, the coupling between the dispersion and gain is shown in Figs. \ref{fig.3}(a)-(d) allows a phase control of the response of the medium. Thus, it is possible that we can turn an absorber at $\beta = 0$ into an amplifier at $\beta = \pi/2$. 

We further investigate the SOP of input at $z = 0$ and the output beam at a distance $z_R$ as shown in Figs. \ref{fig.5}(a)-(i). As stated earlier, achieving the desired polarization state relies on the spatial intensity distribution of both components of the vector beam. Figures \ref{fig.5}(a)-(c) exhibit the initial transverse intensity and polarization distribution of radial, azimuthal, and spiral CV beams, respectively. The intensity distribution has been normalized with its maximum intensity for all figures, enabling us to visualize the diffraction it poses. It is noticed from the Figs. \ref{fig.5}(d)-(f) that under the condition of $\beta = 0$, the change in the polarization state is a consequence of the difference between Re[$\chi_{41}$] and Re[$\chi_{32}$]. We also observed that after propagating a length $z_R$ inside the medium, the radial CV beam was transformed into azimuthal and vice versa. However, the clockwise spiral CV beam changed to an anticlockwise spiral. This phenomenon arises because at each point in the transverse plane, the polarization rotates $\pi/2$. So, it is certain that the second term in Eq. (\ref{eq.22}) dominates. On the other hand, when $\beta = \pi/2$, the polarization rotation is smaller than in the former case, although there is a variation in ellipticity as shown in Figs. \ref{fig.5}(g)-(i). For the case of radial and spiral CV beams, the ellipticity decreases and forms linear polarization along the transverse plane. Meanwhile, we also notice a variation of ellipticity along the radial direction for azimuthal VB. This variation of ellipticity can be connected to Eqs. (\ref{eq.23}) and (\ref{eq.24}). As can be seen from Eq. (\ref{eq.24}) for $z \neq 0$, the variation of ellipticity is a function of spatial coordinates. Therefore, we showed how the polarization and ellipticity change due to the choice of the phase shift among the input fields.
\subsection{Focusing of CV beams}
In this section, we discuss the focusing of CV beams that result from the
%%%%%%%%%%%%%%%%%%%%%%%%%%%%%%%%%%%%%%%%%%%
%				Figure 7
%%%%%%%%%%%%%%%%%%%%%%%%%%%%%%%%%%%%%%%%%%%
\begin{figure}[ht]
	\centering
	\includegraphics[width=\linewidth]{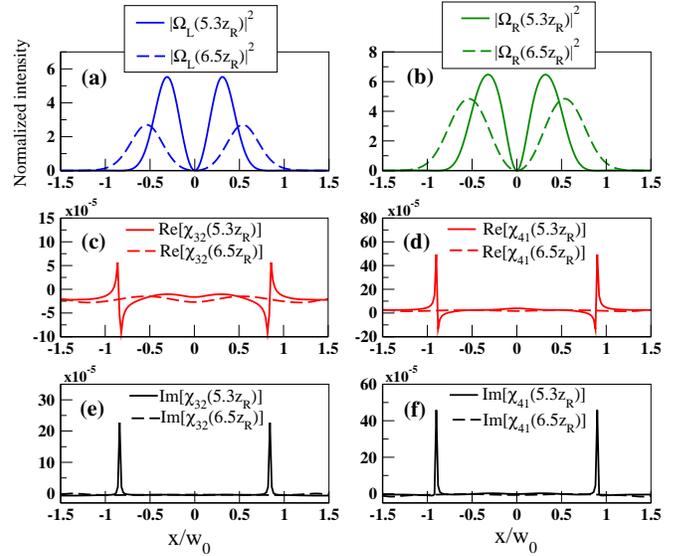}
	\caption{(a)-(b) Illustrate the transverse intensity profile of left and right circular components of azimuthal CV beam (at $z = 5.3z_R$, and $z = 6.5z_R$) respectively which are normalized to the respective peak intensity at $z = 0$. (c)-(d) Shows the Re[$\chi_{32}$], and Re[$\chi_{41}$] for both the two components (at $z = 5.3z_R$, and $z = 6.5z_R$) respectively. (e)-(f) Represent the Im[$\chi_{32}$], and Im[$\chi_{41}$] for both the two components (at $z = 5.3z_R$, and $z = 6.5z_R$) respectively. Parameters remain same as Fig. \ref{fig.6}.}
	\label{fig.7}
\end{figure} 
%%%%%%%%%%%%%%%%%%%%%%%%%%%%%%%%%%%%%%%%%%% 
%%%%%%%%%%%%%%%%%%%%%%%%%%%%%%%%%%%%%%%%%%%
nonlinear effect of the medium. The calculation of coherences, as stated earlier, is no longer valid under weak probe approximation at the strong intensity of CV beams. Therefore, a full numerical solution is desirable. Figures \ref{fig.6}(a)-(c) illustrate the longitudinal profile of radial, azimuthal, and spiral CV beams, respectively. At strong probe intensities, a phenomenon referred to as ``self-focusing" emerges due to the dominant character of the medium's third-order nonlinearity. The white vertical dotted line in Figs. \ref{fig.6}(a)-(c) indicates the lengths at which the first focal spots for radial, azimuthal, and spiral CV beams are formed. These lengths are 2.1$z_R$, 5.3$z_R$ and 2.7$z_R$ for radial, azimuthal and spiral CV beams, respectively. However, the distance of the focal spots may vary depending on the detunings $\Delta_1$ and $\Delta_2$. These detunings are the controlling parameters for the refractive index with optimal gain experienced by the beam propagating through the medium. As the beam propagates through the medium, its intensity increases at the consecutive focal points. Therefore, the beams get distorted at a larger distance due to high gain in the medium. This unique chain-like focusing pattern holds significance in optical trap chains in an optical trap system. To comprehend the mechanism behind such focusing, we have plotted the spatial susceptibilities of azimuthal CV beams for two different distances, as shown in Figs. \ref{fig.7}(c)-(f). Figures \ref{fig.7}(a) and \ref{fig.7}(b) depict the intensity distribution of the left and right circularly polarized components of the azimuthal CV beam at the first focusing and defocusing point marked by the dashed vertical lines (white and green respectively) in Fig. \ref{fig.6}(b). The intensities are normalized with the peak intensities at the input for both probe components. We have observed that both components of the probe VB have been first focused at a distance of $z = 5.3z_R$ and defocused at $z = 6.5z_R$. This phenomenon of focusing and defocusing can be elucidated by referring to Figs. \ref{fig.7}(c)-(f). Figures \ref{fig.7}(c) and \ref{fig.7}(d), delineate that the Re[$\chi_{32}$] and  Re[$\chi_{41}$] for both probe components display higher values at the central region around $x = 0$ for  $5.3z_R$ compared to $6.5z_R$. This small difference in the refractive index is adequate to cause beam focusing and defocusing.  Additionally, in Figs. \ref{fig.7}(e)-(f), the small negative value of Im[$\chi_{32}$], and  Im[$\chi_{41}$] around $x = 0$ resembles gain. This explanation for beam focusing and defocusing also applies to radial and spiral CV beams. Figures \ref{fig.8}(a)-(c) demonstrate the SOP at the first focal point for the radial, azimuthal, and spiral CV beam; 
%%%%%%%%%%%%%%%%%%%%%%%%%%%%%%%%%%%%%%%%%%%
%%%%%%%%%%%%%%%%%%%%%%%%%%%%%%%%%%%%%%%%%%%
%				Figure 8
%%%%%%%%%%%%%%%%%%%%%%%%%%%%%%%%%%%%%%%%%%%
\begin{figure}[ht]
	\centering
	\includegraphics[width=\linewidth]{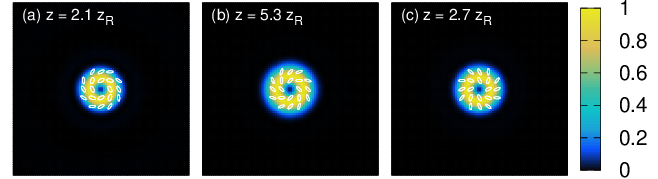}
	\caption{(a)-(c) Transverse intensity and polarization distribution at the first focal point for radial, azimuthal, and spiral CV beams respectively. The intensity is normalized by its maximum intensity for each VB. Parameters remain same as Fig. \ref{fig.6}.}
	\label{fig.8}
\end{figure}
%%%%%%%%%%%%%%%%%%%%%%%%%%%%%%%%%%%%%%%%%%%
respectively. It is observed that the focused beams at different 
%%%%%%%%%%%%%%%%%%%%%%%%%%%%%%%%%%%%%%%%%%%
%				Figure 9
%%%%%%%%%%%%%%%%%%%%%%%%%%%%%%%%%%%%%%%%%%%
\begin{figure}[ht]
	\centering
	\includegraphics[width=\linewidth]{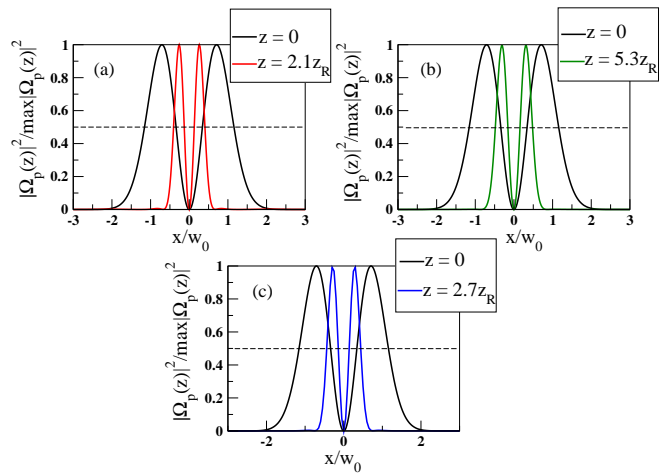}
	\caption{(a)-(c) Transverse intensity profile of radial, azimuthal, and spiral CV beams respectively at $z = 0$ and at first focal point. The black dotted line corresponds to half of the maximum intensity. Parameters remain same as Fig. \ref{fig.6}.}
	\label{fig.9}
\end{figure}
%%%%%%%%%%%%%%%%%%%%%%%%%%%%%%%%%%%%%%%%%%%
propagation distances experience polarization rotation. The three types of CV beams all transform a spiral, resulting in a change in handedness for the initial spiral beam. This observed polarization rotation occurs due to the refractive index disparity between two components as described in Eq. (\ref{eq.22}). To visualize the beam narrowing, all figures in Figs. \ref{fig.8}(a)-(c) have been normalized with their respective maximum peak intensity. It is evident that the spot size of the first focal spot varies for the radial, azimuthal and spiral CV beams. Figures. \ref{fig.9}(a)-(c) illustrate the spot size of the radial, azimuthal, and spiral CV beams at the input and their respective first focal points. As the consecutive focal points have almost the same beam narrowing for each type of beam, we have chosen to measure the spot size only at the first focal point for all three CV beams. At $z = 0$, the spot size at full width half maxima (FWHM) of the central dark region is 0.68$w_0$ for the input beam. Upon focusing, the spot size of the radial, azimuthal, and spiral CV beams have reduced to 0.28$w_0$, 0.32$w_0$, and 0.30$w_0$, respectively, at FWHM. Therefore, the suitable choice of intensity and detunings are indispensable to induce a suitable refractive index in the medium to achieve nonlinear focusing.
\section{CONCLUSION}
In summary, we have developed a theoretical framework to manipulate the susceptibility of a medium to achieve gain or loss depending on the phase shift between a probe beam and a control field. Additionally, we have observed self-focusing of different CV beams at higher densities of the medium and higher intensities of the probe and control fields in a non-degenerate four-level atomic system. The system is characterized by coupled transitions via the left and right circularly polarized components of the PVB and the $\pi$ polarized control field. By adjusting the phase shift between the probe beam and the control field, we can modify the susceptibility of the medium to exhibit loss at $\beta = 0$ and gain at $\beta = \pi/2$. We have also studied the polarization rotation for these two cases and found that for $\beta = 0$, the polarization ellipse rotates 90 degrees at each point in the transverse plane after propagating one Rayleigh length. Moreover, for $\beta = \pi/2$, we notice a change in ellipticity is accompanied by polarization rotation. Furthermore, we investigated the medium response when the probe beam intensity became comparable to the control field intensity. At higher atomic density, we have observed self-focusing behaviour for all three types of CV beams due to the emergence of third-order nonlinearity. Upon propagation by a couple of Rayleigh lengths, the beams exhibit a chain-like self-focusing and defocusing pattern. The SOP is described at the first focal point of each beam. The application of self-focusing has resulted in a reduction in the spot size of the beams, which could prove beneficial in enhancing resolution. Additionally, the chain-like pattern observed may hold significance in the realm of optical trap systems.
\begin{acknowledgments}
T.N.D. gratefully acknowledges funding by Science and Engineering Research Board (Grant No. CRG/2023/001318). 
\end{acknowledgments}
\newpage
\onecolumngrid
%\rule{\textwidth}{0.4pt}
\vspace{1cm}
\appendix*
%\newpage
\section{}
The explicit expressions for $A-H$ are as follows
\begin{subequations}\label{eq.A6}
\begin{align}
	A =& \Gamma_{12} \Gamma_{14} \Gamma_{24} \left(\Gamma_{13}+\Gamma_{31}\right) \Gamma_{34} \Gamma_{42},\\
	B =& \left| \Omega_c\right|^2\big[\Gamma_{24} \left(\Gamma_{13}+\Gamma_{31}\right) \left(\Gamma_{12}+\Gamma_{34}\right) \Gamma_{42}-\Gamma_{14} \left(\Gamma_{24} \Gamma_{31} \Gamma_{34}+\left(\Gamma_{12} \left(\Gamma_{13}+\Gamma_{31}\right)+\Gamma_{31} \Gamma_{34}\right) \Gamma_{42}\right)\big], \\
	C =& \left(\Gamma_{14}+\Gamma_{32}\right) \left(\Gamma_{12}+\Gamma_{34}\right) \left| \Omega_c\right|^2+\Gamma_{12} \Gamma_{14} \Gamma_{32} \Gamma_{34},\\
	D =& -4 \left(\Gamma_{13}+\Gamma_{31}\right) \left(\Gamma_{24}+\Gamma_{42}\right) \left| \Omega_c\right|^2+\Gamma_{24} \Gamma_{31} \Gamma_{42}+\Gamma_{13} \left(\Gamma_{24} \Gamma_{31}+\left(\Gamma_{24}+\Gamma_{31}\right) \Gamma_{42}\right),\\
	E =& \Omega_c^2\big[\Gamma_{12} \Gamma_{13} \left(\Gamma_{14}+\Gamma_{31}\right) \left(\Gamma_{24}+\Gamma_{42}\right)+\Gamma_{34} \left(\Gamma_{14} \Gamma_{24} \Gamma_{31}+\Gamma_{13} \left(\Gamma_{24} \left(\Gamma_{14}+\Gamma_{31}\right)+\Gamma_{31} \Gamma_{42}\right)\right)\big],\\
	F =& \Gamma_{13} \Gamma_{21} \Gamma_{23} \Gamma_{31} \left(\Gamma_{24}+\Gamma_{42}\right) \Gamma_{43},\\
	G =& \left| \Omega_c\right|^2\big[\Gamma_{13} \left(\Gamma_{21} \Gamma_{31} \left(\Gamma_{24}+\Gamma_{42}\right)+\left(\Gamma_{31} \left(\Gamma_{24}+\Gamma_{42}\right)-\Gamma_{23} \Gamma_{42}\right) \Gamma_{43}\right)-\Gamma_{23} \Gamma_{31} (\Gamma_{21} \left(\Gamma_{24}+\Gamma_{42}\right)\nonumber\\
	&+\Gamma_{42} \Gamma_{43})\big], \\
	H =& \Omega_c^2\big[\Gamma_{21} \Gamma_{24} \left(\Gamma_{13}+\Gamma_{31}\right) \left(\Gamma_{23}+\Gamma_{42}\right)+\left(\Gamma_{24} \Gamma_{31} \Gamma_{42}+\Gamma_{13} \left(\Gamma_{24} \Gamma_{42}+\Gamma_{23} \left(\Gamma_{24}+\Gamma_{42}\right)\right)\right) \Gamma_{43}\big]. 
\end{align}
\end{subequations} 
In Eqs. (\ref{eq.A6}) the gammas are defined as
\begin{subequations}
	\begin{align}
		&\Gamma_{31} = \bigg(i\Delta_4 -\dfrac{1}{2}\bigg),\\
		&\Gamma_{32} = \bigg(i(\Delta_2 + \Delta_4 - \Delta_1)-\dfrac{1}{2}\bigg),\\
		&\Gamma_{41} = \bigg(i\Delta_1 - \dfrac{1}{2}\bigg),\\
		&\Gamma_{42} = \bigg(i\Delta_2 - \dfrac{1}{2}\bigg),\\
		&\Gamma_{34} = \bigg(i(\Delta_4-\Delta_1) - 1\bigg),\\
		&\Gamma_{12} = -\bigg(\gamma_{12} + i(\Delta_1-\Delta_2)\bigg),
	\end{align}
\end{subequations}
and the complex conjugates are $\Gamma_{13} = \Gamma_{31}^*,\Gamma_{23} = \Gamma_{32}^*,\Gamma_{14} = \Gamma_{41}^*,\Gamma_{24} = \Gamma_{42}^*,\Gamma_{43} = \Gamma_{34}^*,\Gamma_{21} = \Gamma_{12}^*$. The decoherence rate between two ground states is denoted by $\gamma_{12}$. At VB intensities, comparable to the control field intensity, the coherences calculated in perturbation method becomes invalid. To account for the impact of high VB intensity on polarization rotation and beam propagation, the pertinent coherences are computed numerically using Gaussian elimination.
\newpage
\twocolumngrid
\bibliography{paper}% Produces the bibliography via BibTeX.

\begin{thebibliography}{47}%
\makeatletter
\providecommand \@ifxundefined [1]{%
 \@ifx{#1\undefined}
}%
\providecommand \@ifnum [1]{%
 \ifnum #1\expandafter \@firstoftwo
 \else \expandafter \@secondoftwo
 \fi
}%
\providecommand \@ifx [1]{%
 \ifx #1\expandafter \@firstoftwo
 \else \expandafter \@secondoftwo
 \fi
}%
\providecommand \natexlab [1]{#1}%
\providecommand \enquote  [1]{``#1''}%
\providecommand \bibnamefont  [1]{#1}%
\providecommand \bibfnamefont [1]{#1}%
\providecommand \citenamefont [1]{#1}%
\providecommand \href@noop [0]{\@secondoftwo}%
\providecommand \href [0]{\begingroup \@sanitize@url \@href}%
\providecommand \@href[1]{\@@startlink{#1}\@@href}%
\providecommand \@@href[1]{\endgroup#1\@@endlink}%
\providecommand \@sanitize@url [0]{\catcode `\\12\catcode `\$12\catcode
  `\&12\catcode `\#12\catcode `\^12\catcode `\_12\catcode `\%12\relax}%
\providecommand \@@startlink[1]{}%
\providecommand \@@endlink[0]{}%
\providecommand \url  [0]{\begingroup\@sanitize@url \@url }%
\providecommand \@url [1]{\endgroup\@href {#1}{\urlprefix }}%
\providecommand \urlprefix  [0]{URL }%
\providecommand \Eprint [0]{\href }%
\providecommand \doibase [0]{https://doi.org/}%
\providecommand \selectlanguage [0]{\@gobble}%
\providecommand \bibinfo  [0]{\@secondoftwo}%
\providecommand \bibfield  [0]{\@secondoftwo}%
\providecommand \translation [1]{[#1]}%
\providecommand \BibitemOpen [0]{}%
\providecommand \bibitemStop [0]{}%
\providecommand \bibitemNoStop [0]{.\EOS\space}%
\providecommand \EOS [0]{\spacefactor3000\relax}%
\providecommand \BibitemShut  [1]{\csname bibitem#1\endcsname}%
\let\auto@bib@innerbib\@empty
%</preamble>
\bibitem [{\citenamefont {Fleischhauer}\ and\ \citenamefont
  {Lukin}(2002)}]{intro1}%
  \BibitemOpen
  \bibfield  {author} {\bibinfo {author} {\bibfnamefont {M.}~\bibnamefont
  {Fleischhauer}}\ and\ \bibinfo {author} {\bibfnamefont {M.~D.}\ \bibnamefont
  {Lukin}},\ }\bibfield  {title} {\bibinfo {title} {Quantum memory for photons:
  Dark-state polaritons},\ }\href {https://doi.org/10.1103/PhysRevA.65.022314}
  {\bibfield  {journal} {\bibinfo  {journal} {Phys. Rev. A}\ }\textbf {\bibinfo
  {volume} {65}},\ \bibinfo {pages} {022314} (\bibinfo {year}
  {2002})}\BibitemShut {NoStop}%
\bibitem [{\citenamefont {Glauber}(1963)}]{intro2}%
  \BibitemOpen
  \bibfield  {author} {\bibinfo {author} {\bibfnamefont {R.~J.}\ \bibnamefont
  {Glauber}},\ }\bibfield  {title} {\bibinfo {title} {The quantum theory of
  optical coherence},\ }\href {https://doi.org/10.1103/PhysRev.130.2529}
  {\bibfield  {journal} {\bibinfo  {journal} {Phys. Rev.}\ }\textbf {\bibinfo
  {volume} {130}},\ \bibinfo {pages} {2529} (\bibinfo {year}
  {1963})}\BibitemShut {NoStop}%
\bibitem [{\citenamefont {Durnin}(1987)}]{intro3}%
  \BibitemOpen
  \bibfield  {author} {\bibinfo {author} {\bibfnamefont {J.}~\bibnamefont
  {Durnin}},\ }\bibfield  {title} {\bibinfo {title} {Exact solutions for
  nondiffracting beams. i. the scalar theory},\ }\href
  {https://doi.org/10.1364/JOSAA.4.000651} {\bibfield  {journal} {\bibinfo
  {journal} {J. Opt. Soc. Am. A}\ }\textbf {\bibinfo {volume} {4}},\ \bibinfo
  {pages} {651} (\bibinfo {year} {1987})}\BibitemShut {NoStop}%
\bibitem [{\citenamefont {Erikson}\ and\ \citenamefont {Singh}(1994)}]{intro4}%
  \BibitemOpen
  \bibfield  {author} {\bibinfo {author} {\bibfnamefont {W.~L.}\ \bibnamefont
  {Erikson}}\ and\ \bibinfo {author} {\bibfnamefont {S.}~\bibnamefont
  {Singh}},\ }\bibfield  {title} {\bibinfo {title} {Polarization properties of
  maxwell-gaussian laser beams},\ }\href
  {https://doi.org/10.1103/PhysRevE.49.5778} {\bibfield  {journal} {\bibinfo
  {journal} {Phys. Rev. E}\ }\textbf {\bibinfo {volume} {49}},\ \bibinfo
  {pages} {5778} (\bibinfo {year} {1994})}\BibitemShut {NoStop}%
\bibitem [{\citenamefont {Korotkova}\ \emph {et~al.}(2005)\citenamefont
  {Korotkova}, \citenamefont {Hoover}, \citenamefont {Gamiz},\ and\
  \citenamefont {Wolf}}]{intro5}%
  \BibitemOpen
  \bibfield  {author} {\bibinfo {author} {\bibfnamefont {O.}~\bibnamefont
  {Korotkova}}, \bibinfo {author} {\bibfnamefont {B.~G.}\ \bibnamefont
  {Hoover}}, \bibinfo {author} {\bibfnamefont {V.~L.}\ \bibnamefont {Gamiz}},\
  and\ \bibinfo {author} {\bibfnamefont {E.}~\bibnamefont {Wolf}},\ }\bibfield
  {title} {\bibinfo {title} {Coherence and polarization properties of far
  fields generated by quasi-homogeneous planar electromagnetic sources},\
  }\href {https://doi.org/10.1364/JOSAA.22.002547} {\bibfield  {journal}
  {\bibinfo  {journal} {J. Opt. Soc. Am. A}\ }\textbf {\bibinfo {volume}
  {22}},\ \bibinfo {pages} {2547} (\bibinfo {year} {2005})}\BibitemShut
  {NoStop}%
\bibitem [{\citenamefont {Deng}\ \emph {et~al.}(2008)\citenamefont {Deng},
  \citenamefont {Yu}, \citenamefont {Xu}, \citenamefont {Shao},\ and\
  \citenamefont {Fan}}]{intro6}%
  \BibitemOpen
  \bibfield  {author} {\bibinfo {author} {\bibfnamefont {D.}~\bibnamefont
  {Deng}}, \bibinfo {author} {\bibfnamefont {H.}~\bibnamefont {Yu}}, \bibinfo
  {author} {\bibfnamefont {S.}~\bibnamefont {Xu}}, \bibinfo {author}
  {\bibfnamefont {J.}~\bibnamefont {Shao}},\ and\ \bibinfo {author}
  {\bibfnamefont {Z.}~\bibnamefont {Fan}},\ }\bibfield  {title} {\bibinfo
  {title} {Propagation and polarization properties of hollow gaussian beams in
  uniaxial crystals},\ }\href
  {https://doi.org/https://doi.org/10.1016/j.optcom.2007.09.038} {\bibfield
  {journal} {\bibinfo  {journal} {Optics Communications}\ }\textbf {\bibinfo
  {volume} {281}},\ \bibinfo {pages} {202} (\bibinfo {year}
  {2008})}\BibitemShut {NoStop}%
\bibitem [{\citenamefont {Ficek}\ and\ \citenamefont {Swain}(2005)}]{intro7}%
  \BibitemOpen
  \bibfield  {author} {\bibinfo {author} {\bibfnamefont {Z.}~\bibnamefont
  {Ficek}}\ and\ \bibinfo {author} {\bibfnamefont {S.}~\bibnamefont {Swain}},\
  }\href@noop {} {\emph {\bibinfo {title} {Quantum interference and coherence:
  theory and experiments}}},\ Vol.\ \bibinfo {volume} {100}\ (\bibinfo
  {publisher} {Springer Science \& Business Media},\ \bibinfo {year}
  {2005})\BibitemShut {NoStop}%
\bibitem [{\citenamefont {Kiffner}\ \emph {et~al.}(2006)\citenamefont
  {Kiffner}, \citenamefont {Evers},\ and\ \citenamefont {Keitel}}]{intro8}%
  \BibitemOpen
  \bibfield  {author} {\bibinfo {author} {\bibfnamefont {M.}~\bibnamefont
  {Kiffner}}, \bibinfo {author} {\bibfnamefont {J.}~\bibnamefont {Evers}},\
  and\ \bibinfo {author} {\bibfnamefont {C.~H.}\ \bibnamefont {Keitel}},\
  }\bibfield  {title} {\bibinfo {title} {Interference in the resonance
  fluorescence of two incoherently coupled transitions},\ }\href
  {https://doi.org/10.1103/PhysRevA.73.063814} {\bibfield  {journal} {\bibinfo
  {journal} {Phys. Rev. A}\ }\textbf {\bibinfo {volume} {73}},\ \bibinfo
  {pages} {063814} (\bibinfo {year} {2006})}\BibitemShut {NoStop}%
\bibitem [{\citenamefont {Delagnes}\ and\ \citenamefont
  {Bouchene}(2007)}]{intro9}%
  \BibitemOpen
  \bibfield  {author} {\bibinfo {author} {\bibfnamefont {J.~C.}\ \bibnamefont
  {Delagnes}}\ and\ \bibinfo {author} {\bibfnamefont {M.~A.}\ \bibnamefont
  {Bouchene}},\ }\bibfield  {title} {\bibinfo {title} {Coherent control of
  light shifts in an atomic system: Modulation of the medium gain},\ }\href
  {https://doi.org/10.1103/PhysRevLett.98.053602} {\bibfield  {journal}
  {\bibinfo  {journal} {Phys. Rev. Lett.}\ }\textbf {\bibinfo {volume} {98}},\
  \bibinfo {pages} {053602} (\bibinfo {year} {2007})}\BibitemShut {NoStop}%
\bibitem [{\citenamefont {Hashmi}\ and\ \citenamefont
  {Bouchene}(2008{\natexlab{a}})}]{intro10}%
  \BibitemOpen
  \bibfield  {author} {\bibinfo {author} {\bibfnamefont {F.~A.}\ \bibnamefont
  {Hashmi}}\ and\ \bibinfo {author} {\bibfnamefont {M.~A.}\ \bibnamefont
  {Bouchene}},\ }\bibfield  {title} {\bibinfo {title} {Slowing light through
  zeeman coherence oscillations in a duplicated two-level system},\ }\href
  {https://doi.org/10.1103/PhysRevA.77.051803} {\bibfield  {journal} {\bibinfo
  {journal} {Phys. Rev. A}\ }\textbf {\bibinfo {volume} {77}},\ \bibinfo
  {pages} {051803(R)} (\bibinfo {year} {2008}{\natexlab{a}})}\BibitemShut
  {NoStop}%
\bibitem [{\citenamefont {Jin}\ \emph {et~al.}(2011)\citenamefont {Jin},
  \citenamefont {Niu},\ and\ \citenamefont {Gong}}]{intro11}%
  \BibitemOpen
  \bibfield  {author} {\bibinfo {author} {\bibfnamefont {L.}~\bibnamefont
  {Jin}}, \bibinfo {author} {\bibfnamefont {Y.}~\bibnamefont {Niu}},\ and\
  \bibinfo {author} {\bibfnamefont {S.}~\bibnamefont {Gong}},\ }\bibfield
  {title} {\bibinfo {title} {Phase control of spatial interference from two
  duplicated two-level atoms},\ }\href
  {https://doi.org/10.1103/PhysRevA.83.023410} {\bibfield  {journal} {\bibinfo
  {journal} {Phys. Rev. A}\ }\textbf {\bibinfo {volume} {83}},\ \bibinfo
  {pages} {023410} (\bibinfo {year} {2011})}\BibitemShut {NoStop}%
\bibitem [{\citenamefont {Hashmi}\ and\ \citenamefont
  {Bouchene}(2008{\natexlab{b}})}]{intro12}%
  \BibitemOpen
  \bibfield  {author} {\bibinfo {author} {\bibfnamefont {F.~A.}\ \bibnamefont
  {Hashmi}}\ and\ \bibinfo {author} {\bibfnamefont {M.~A.}\ \bibnamefont
  {Bouchene}},\ }\bibfield  {title} {\bibinfo {title} {Coherent control of the
  effective susceptibility through wave mixing in a duplicated two-level
  system},\ }\href {https://doi.org/10.1103/PhysRevLett.101.213601} {\bibfield
  {journal} {\bibinfo  {journal} {Phys. Rev. Lett.}\ }\textbf {\bibinfo
  {volume} {101}},\ \bibinfo {pages} {213601} (\bibinfo {year}
  {2008}{\natexlab{b}})}\BibitemShut {NoStop}%
\bibitem [{\citenamefont {Wu}\ \emph {et~al.}(2010)\citenamefont {Wu},
  \citenamefont {Lü},\ and\ \citenamefont {Zheng}}]{intro13}%
  \BibitemOpen
  \bibfield  {author} {\bibinfo {author} {\bibfnamefont {J.}~\bibnamefont
  {Wu}}, \bibinfo {author} {\bibfnamefont {X.-Y.}\ \bibnamefont {Lü}},\ and\
  \bibinfo {author} {\bibfnamefont {L.-L.}\ \bibnamefont {Zheng}},\ }\bibfield
  {title} {\bibinfo {title} {Controllable optical bistability and
  multistability in a double two-level atomic system},\ }\href
  {https://doi.org/10.1088/0953-4075/43/16/161003} {\bibfield  {journal}
  {\bibinfo  {journal} {J. Phys. B: At., Mol. Opt. Phys.}\ }\textbf {\bibinfo
  {volume} {43}},\ \bibinfo {pages} {161003} (\bibinfo {year}
  {2010})}\BibitemShut {NoStop}%
\bibitem [{\citenamefont {Zhu}\ \emph {et~al.}(2016)\citenamefont {Zhu},
  \citenamefont {Yang}, \citenamefont {Xie}, \citenamefont {Liu}, \citenamefont
  {Liu},\ and\ \citenamefont {Lee}}]{intro14}%
  \BibitemOpen
  \bibfield  {author} {\bibinfo {author} {\bibfnamefont {Z.}~\bibnamefont
  {Zhu}}, \bibinfo {author} {\bibfnamefont {W.-X.}\ \bibnamefont {Yang}},
  \bibinfo {author} {\bibfnamefont {X.-T.}\ \bibnamefont {Xie}}, \bibinfo
  {author} {\bibfnamefont {S.}~\bibnamefont {Liu}}, \bibinfo {author}
  {\bibfnamefont {S.}~\bibnamefont {Liu}},\ and\ \bibinfo {author}
  {\bibfnamefont {R.-K.}\ \bibnamefont {Lee}},\ }\bibfield  {title} {\bibinfo
  {title} {Three-dimensional atom localization from spatial interference in a
  double two-level atomic system},\ }\href
  {https://doi.org/10.1103/PhysRevA.94.013826} {\bibfield  {journal} {\bibinfo
  {journal} {Phys. Rev. A}\ }\textbf {\bibinfo {volume} {94}},\ \bibinfo
  {pages} {013826} (\bibinfo {year} {2016})}\BibitemShut {NoStop}%
\bibitem [{\citenamefont {Kani}\ and\ \citenamefont {Wanare}(2018)}]{intro15}%
  \BibitemOpen
  \bibfield  {author} {\bibinfo {author} {\bibfnamefont {A.}~\bibnamefont
  {Kani}}\ and\ \bibinfo {author} {\bibfnamefont {H.}~\bibnamefont {Wanare}},\
  }\bibfield  {title} {\bibinfo {title} {Anisotropic nonlinear optics based on
  quantum interference},\ }\href {https://doi.org/10.1209/0295-5075/120/33001}
  {\bibfield  {journal} {\bibinfo  {journal} {Europhysics Letters}\ }\textbf
  {\bibinfo {volume} {120}},\ \bibinfo {pages} {33001} (\bibinfo {year}
  {2018})}\BibitemShut {NoStop}%
\bibitem [{\citenamefont {Zhan}(2009)}]{intro16}%
  \BibitemOpen
  \bibfield  {author} {\bibinfo {author} {\bibfnamefont {Q.}~\bibnamefont
  {Zhan}},\ }\bibfield  {title} {\bibinfo {title} {Cylindrical vector beams:
  from mathematical concepts to applications},\ }\href
  {https://doi.org/10.1364/AOP.1.000001} {\bibfield  {journal} {\bibinfo
  {journal} {Adv. Opt. Photon.}\ }\textbf {\bibinfo {volume} {1}},\ \bibinfo
  {pages} {1} (\bibinfo {year} {2009})}\BibitemShut {NoStop}%
\bibitem [{\citenamefont {Beckley}\ \emph {et~al.}(2010)\citenamefont
  {Beckley}, \citenamefont {Brown},\ and\ \citenamefont {Alonso}}]{intro17}%
  \BibitemOpen
  \bibfield  {author} {\bibinfo {author} {\bibfnamefont {A.~M.}\ \bibnamefont
  {Beckley}}, \bibinfo {author} {\bibfnamefont {T.~G.}\ \bibnamefont {Brown}},\
  and\ \bibinfo {author} {\bibfnamefont {M.~A.}\ \bibnamefont {Alonso}},\
  }\bibfield  {title} {\bibinfo {title} {Full poincar\'{e} beams},\ }\href
  {https://doi.org/10.1364/OE.18.010777} {\bibfield  {journal} {\bibinfo
  {journal} {Opt. Express}\ }\textbf {\bibinfo {volume} {18}},\ \bibinfo
  {pages} {10777} (\bibinfo {year} {2010})}\BibitemShut {NoStop}%
\bibitem [{\citenamefont {Galvez}\ \emph {et~al.}(2012)\citenamefont {Galvez},
  \citenamefont {Khadka}, \citenamefont {Schubert},\ and\ \citenamefont
  {Nomoto}}]{intro18}%
  \BibitemOpen
  \bibfield  {author} {\bibinfo {author} {\bibfnamefont {E.~J.}\ \bibnamefont
  {Galvez}}, \bibinfo {author} {\bibfnamefont {S.}~\bibnamefont {Khadka}},
  \bibinfo {author} {\bibfnamefont {W.~H.}\ \bibnamefont {Schubert}},\ and\
  \bibinfo {author} {\bibfnamefont {S.}~\bibnamefont {Nomoto}},\ }\bibfield
  {title} {\bibinfo {title} {Poincar\'{e}-beam patterns produced by
  nonseparable superpositions of laguerre--gauss and polarization modes of
  light},\ }\href {https://doi.org/10.1364/AO.51.002925} {\bibfield  {journal}
  {\bibinfo  {journal} {Appl. Opt.}\ }\textbf {\bibinfo {volume} {51}},\
  \bibinfo {pages} {2925} (\bibinfo {year} {2012})}\BibitemShut {NoStop}%
\bibitem [{\citenamefont {Yao}\ and\ \citenamefont {Padgett}(2011)}]{intro19}%
  \BibitemOpen
  \bibfield  {author} {\bibinfo {author} {\bibfnamefont {A.~M.}\ \bibnamefont
  {Yao}}\ and\ \bibinfo {author} {\bibfnamefont {M.~J.}\ \bibnamefont
  {Padgett}},\ }\bibfield  {title} {\bibinfo {title} {Orbital angular momentum:
  origins, behavior and applications},\ }\href
  {https://doi.org/10.1364/AOP.3.000161} {\bibfield  {journal} {\bibinfo
  {journal} {Adv. Opt. Photon.}\ }\textbf {\bibinfo {volume} {3}},\ \bibinfo
  {pages} {161} (\bibinfo {year} {2011})}\BibitemShut {NoStop}%
\bibitem [{\citenamefont {Waller}\ and\ \citenamefont {von
  Freymann}(2013)}]{intro20}%
  \BibitemOpen
  \bibfield  {author} {\bibinfo {author} {\bibfnamefont {E.~H.}\ \bibnamefont
  {Waller}}\ and\ \bibinfo {author} {\bibfnamefont {G.}~\bibnamefont {von
  Freymann}},\ }\bibfield  {title} {\bibinfo {title} {Independent spatial
  intensity, phase and polarization distributions},\ }\href
  {https://doi.org/10.1364/OE.21.028167} {\bibfield  {journal} {\bibinfo
  {journal} {Opt. Express}\ }\textbf {\bibinfo {volume} {21}},\ \bibinfo
  {pages} {28167} (\bibinfo {year} {2013})}\BibitemShut {NoStop}%
\bibitem [{\citenamefont {Quabis}\ \emph {et~al.}(2000)\citenamefont {Quabis},
  \citenamefont {Dorn}, \citenamefont {Eberler}, \citenamefont {Gl{\"o}ckl},\
  and\ \citenamefont {Leuchs}}]{intro21}%
  \BibitemOpen
  \bibfield  {author} {\bibinfo {author} {\bibfnamefont {S.}~\bibnamefont
  {Quabis}}, \bibinfo {author} {\bibfnamefont {R.}~\bibnamefont {Dorn}},
  \bibinfo {author} {\bibfnamefont {M.}~\bibnamefont {Eberler}}, \bibinfo
  {author} {\bibfnamefont {O.}~\bibnamefont {Gl{\"o}ckl}},\ and\ \bibinfo
  {author} {\bibfnamefont {G.}~\bibnamefont {Leuchs}},\ }\bibfield  {title}
  {\bibinfo {title} {Focusing light to a tighter spot},\ }\href
  {https://www.sciencedirect.com/science/article/pii/S0030401899007294}
  {\bibfield  {journal} {\bibinfo  {journal} {Optics Communications}\ }\textbf
  {\bibinfo {volume} {179}},\ \bibinfo {pages} {1} (\bibinfo {year}
  {2000})}\BibitemShut {NoStop}%
\bibitem [{\citenamefont {Dorn}\ \emph {et~al.}(2003)\citenamefont {Dorn},
  \citenamefont {Quabis},\ and\ \citenamefont {Leuchs}}]{intro22}%
  \BibitemOpen
  \bibfield  {author} {\bibinfo {author} {\bibfnamefont {R.}~\bibnamefont
  {Dorn}}, \bibinfo {author} {\bibfnamefont {S.}~\bibnamefont {Quabis}},\ and\
  \bibinfo {author} {\bibfnamefont {G.}~\bibnamefont {Leuchs}},\ }\bibfield
  {title} {\bibinfo {title} {Sharper focus for a radially polarized light
  beam},\ }\href {https://doi.org/10.1103/PhysRevLett.91.233901} {\bibfield
  {journal} {\bibinfo  {journal} {Phys. Rev. Lett.}\ }\textbf {\bibinfo
  {volume} {91}},\ \bibinfo {pages} {233901} (\bibinfo {year}
  {2003})}\BibitemShut {NoStop}%
\bibitem [{\citenamefont {Jia}\ \emph {et~al.}(2005)\citenamefont {Jia},
  \citenamefont {Gan},\ and\ \citenamefont {Gu}}]{intro23}%
  \BibitemOpen
  \bibfield  {author} {\bibinfo {author} {\bibfnamefont {B.}~\bibnamefont
  {Jia}}, \bibinfo {author} {\bibfnamefont {X.}~\bibnamefont {Gan}},\ and\
  \bibinfo {author} {\bibfnamefont {M.}~\bibnamefont {Gu}},\ }\bibfield
  {title} {\bibinfo {title} {Direct measurement of a radially polarized focused
  evanescent field facilitated by a single lcd},\ }\href
  {https://doi.org/10.1364/OPEX.13.006821} {\bibfield  {journal} {\bibinfo
  {journal} {Opt. Express}\ }\textbf {\bibinfo {volume} {13}},\ \bibinfo
  {pages} {6821} (\bibinfo {year} {2005})}\BibitemShut {NoStop}%
\bibitem [{\citenamefont {Hao}\ and\ \citenamefont {Leger}(2007)}]{intro24}%
  \BibitemOpen
  \bibfield  {author} {\bibinfo {author} {\bibfnamefont {B.}~\bibnamefont
  {Hao}}\ and\ \bibinfo {author} {\bibfnamefont {J.}~\bibnamefont {Leger}},\
  }\bibfield  {title} {\bibinfo {title} {Experimental measurement of
  longitudinal component in the vicinity of focused radially polarized beam},\
  }\href {https://doi.org/10.1364/OE.15.003550} {\bibfield  {journal} {\bibinfo
   {journal} {Opt. Express}\ }\textbf {\bibinfo {volume} {15}},\ \bibinfo
  {pages} {3550} (\bibinfo {year} {2007})}\BibitemShut {NoStop}%
\bibitem [{\citenamefont {Willig}\ \emph {et~al.}(2006)\citenamefont {Willig},
  \citenamefont {Rizzoli}, \citenamefont {Westphal}, \citenamefont {Jahn},\
  and\ \citenamefont {Hell}}]{intro25}%
  \BibitemOpen
  \bibfield  {author} {\bibinfo {author} {\bibfnamefont {K.~I.}\ \bibnamefont
  {Willig}}, \bibinfo {author} {\bibfnamefont {S.~O.}\ \bibnamefont {Rizzoli}},
  \bibinfo {author} {\bibfnamefont {V.}~\bibnamefont {Westphal}}, \bibinfo
  {author} {\bibfnamefont {R.}~\bibnamefont {Jahn}},\ and\ \bibinfo {author}
  {\bibfnamefont {S.~W.}\ \bibnamefont {Hell}},\ }\bibfield  {title} {\bibinfo
  {title} {Sted microscopy reveals that synaptotagmin remains clustered after
  synaptic vesicle exocytosis},\ }\href {https://doi.org/10.1038/nature04592}
  {\bibfield  {journal} {\bibinfo  {journal} {Nature}\ }\textbf {\bibinfo
  {volume} {440}},\ \bibinfo {pages} {935} (\bibinfo {year}
  {2006})}\BibitemShut {NoStop}%
\bibitem [{\citenamefont {T\"{o}r\"{o}k}\ and\ \citenamefont
  {Munro}(2004)}]{intro26}%
  \BibitemOpen
  \bibfield  {author} {\bibinfo {author} {\bibfnamefont {P.}~\bibnamefont
  {T\"{o}r\"{o}k}}\ and\ \bibinfo {author} {\bibfnamefont {P.}~\bibnamefont
  {Munro}},\ }\bibfield  {title} {\bibinfo {title} {The use of gauss-laguerre
  vector beams in sted microscopy},\ }\href
  {https://doi.org/10.1364/OPEX.12.003605} {\bibfield  {journal} {\bibinfo
  {journal} {Opt. Express}\ }\textbf {\bibinfo {volume} {12}},\ \bibinfo
  {pages} {3605} (\bibinfo {year} {2004})}\BibitemShut {NoStop}%
\bibitem [{\citenamefont {Youngworth}\ and\ \citenamefont
  {Brown}(2000)}]{intro27}%
  \BibitemOpen
  \bibfield  {author} {\bibinfo {author} {\bibfnamefont {K.~S.}\ \bibnamefont
  {Youngworth}}\ and\ \bibinfo {author} {\bibfnamefont {T.~G.}\ \bibnamefont
  {Brown}},\ }\bibfield  {title} {\bibinfo {title} {Focusing of high numerical
  aperture cylindrical-vector beams},\ }\href
  {https://doi.org/10.1364/OE.7.000077} {\bibfield  {journal} {\bibinfo
  {journal} {Opt. Express}\ }\textbf {\bibinfo {volume} {7}},\ \bibinfo {pages}
  {77} (\bibinfo {year} {2000})}\BibitemShut {NoStop}%
\bibitem [{\citenamefont {Khonina}\ and\ \citenamefont
  {Golub}(2012)}]{intro28}%
  \BibitemOpen
  \bibfield  {author} {\bibinfo {author} {\bibfnamefont {S.~N.}\ \bibnamefont
  {Khonina}}\ and\ \bibinfo {author} {\bibfnamefont {I.}~\bibnamefont
  {Golub}},\ }\bibfield  {title} {\bibinfo {title} {How low can sted go?
  comparison of different write-erase beam combinations for stimulated emission
  depletion microscopy},\ }\href {https://doi.org/10.1364/JOSAA.29.002242}
  {\bibfield  {journal} {\bibinfo  {journal} {J. Opt. Soc. Am. A}\ }\textbf
  {\bibinfo {volume} {29}},\ \bibinfo {pages} {2242} (\bibinfo {year}
  {2012})}\BibitemShut {NoStop}%
\bibitem [{\citenamefont {Skelton}\ \emph {et~al.}(2013)\citenamefont
  {Skelton}, \citenamefont {Sergides}, \citenamefont {Saija}, \citenamefont
  {Iat\`{i}}, \citenamefont {Marag\'{o}},\ and\ \citenamefont
  {Jones}}]{intro29}%
  \BibitemOpen
  \bibfield  {author} {\bibinfo {author} {\bibfnamefont {S.~E.}\ \bibnamefont
  {Skelton}}, \bibinfo {author} {\bibfnamefont {M.}~\bibnamefont {Sergides}},
  \bibinfo {author} {\bibfnamefont {R.}~\bibnamefont {Saija}}, \bibinfo
  {author} {\bibfnamefont {M.~A.}\ \bibnamefont {Iat\`{i}}}, \bibinfo {author}
  {\bibfnamefont {O.~M.}\ \bibnamefont {Marag\'{o}}},\ and\ \bibinfo {author}
  {\bibfnamefont {P.~H.}\ \bibnamefont {Jones}},\ }\bibfield  {title} {\bibinfo
  {title} {Trapping volume control in optical tweezers using cylindrical vector
  beams},\ }\href {https://doi.org/10.1364/OL.38.000028} {\bibfield  {journal}
  {\bibinfo  {journal} {Opt. Lett.}\ }\textbf {\bibinfo {volume} {38}},\
  \bibinfo {pages} {28} (\bibinfo {year} {2013})}\BibitemShut {NoStop}%
\bibitem [{\citenamefont {Novotny}\ \emph {et~al.}(2001)\citenamefont
  {Novotny}, \citenamefont {Beversluis}, \citenamefont {Youngworth},\ and\
  \citenamefont {Brown}}]{intro30}%
  \BibitemOpen
  \bibfield  {author} {\bibinfo {author} {\bibfnamefont {L.}~\bibnamefont
  {Novotny}}, \bibinfo {author} {\bibfnamefont {M.~R.}\ \bibnamefont
  {Beversluis}}, \bibinfo {author} {\bibfnamefont {K.~S.}\ \bibnamefont
  {Youngworth}},\ and\ \bibinfo {author} {\bibfnamefont {T.~G.}\ \bibnamefont
  {Brown}},\ }\bibfield  {title} {\bibinfo {title} {Longitudinal field modes
  probed by single molecules},\ }\href
  {https://doi.org/10.1103/PhysRevLett.86.5251} {\bibfield  {journal} {\bibinfo
   {journal} {Phys. Rev. Lett.}\ }\textbf {\bibinfo {volume} {86}},\ \bibinfo
  {pages} {5251} (\bibinfo {year} {2001})}\BibitemShut {NoStop}%
\bibitem [{\citenamefont {La~Porta}\ and\ \citenamefont
  {Wang}(2004)}]{intro31}%
  \BibitemOpen
  \bibfield  {author} {\bibinfo {author} {\bibfnamefont {A.}~\bibnamefont
  {La~Porta}}\ and\ \bibinfo {author} {\bibfnamefont {M.~D.}\ \bibnamefont
  {Wang}},\ }\bibfield  {title} {\bibinfo {title} {Optical torque wrench:
  Angular trapping, rotation, and torque detection of quartz microparticles},\
  }\href {https://doi.org/10.1103/PhysRevLett.92.190801} {\bibfield  {journal}
  {\bibinfo  {journal} {Phys. Rev. Lett.}\ }\textbf {\bibinfo {volume} {92}},\
  \bibinfo {pages} {190801} (\bibinfo {year} {2004})}\BibitemShut {NoStop}%
\bibitem [{\citenamefont {Sick}\ \emph {et~al.}(2000)\citenamefont {Sick},
  \citenamefont {Hecht},\ and\ \citenamefont {Novotny}}]{intro32}%
  \BibitemOpen
  \bibfield  {author} {\bibinfo {author} {\bibfnamefont {B.}~\bibnamefont
  {Sick}}, \bibinfo {author} {\bibfnamefont {B.}~\bibnamefont {Hecht}},\ and\
  \bibinfo {author} {\bibfnamefont {L.}~\bibnamefont {Novotny}},\ }\bibfield
  {title} {\bibinfo {title} {Orientational imaging of single molecules by
  annular illumination},\ }\href {https://doi.org/10.1103/PhysRevLett.85.4482}
  {\bibfield  {journal} {\bibinfo  {journal} {Phys. Rev. Lett.}\ }\textbf
  {\bibinfo {volume} {85}},\ \bibinfo {pages} {4482} (\bibinfo {year}
  {2000})}\BibitemShut {NoStop}%
\bibitem [{\citenamefont {Fatemi}(2011)}]{intro33}%
  \BibitemOpen
  \bibfield  {author} {\bibinfo {author} {\bibfnamefont {F.~K.}\ \bibnamefont
  {Fatemi}},\ }\bibfield  {title} {\bibinfo {title} {Cylindrical vector beams
  for rapid polarization-dependent measurements in atomic systems},\ }\href
  {https://doi.org/10.1364/OE.19.025143} {\bibfield  {journal} {\bibinfo
  {journal} {Opt. Express}\ }\textbf {\bibinfo {volume} {19}},\ \bibinfo
  {pages} {25143} (\bibinfo {year} {2011})}\BibitemShut {NoStop}%
\bibitem [{\citenamefont {Milione}\ \emph {et~al.}(2015)\citenamefont
  {Milione}, \citenamefont {Lavery}, \citenamefont {Huang}, \citenamefont
  {Ren}, \citenamefont {Xie}, \citenamefont {Nguyen}, \citenamefont {Karimi},
  \citenamefont {Marrucci}, \citenamefont {Nolan}, \citenamefont {Alfano},\
  and\ \citenamefont {Willner}}]{intro34}%
  \BibitemOpen
  \bibfield  {author} {\bibinfo {author} {\bibfnamefont {G.}~\bibnamefont
  {Milione}}, \bibinfo {author} {\bibfnamefont {M.~P.~J.}\ \bibnamefont
  {Lavery}}, \bibinfo {author} {\bibfnamefont {H.}~\bibnamefont {Huang}},
  \bibinfo {author} {\bibfnamefont {Y.}~\bibnamefont {Ren}}, \bibinfo {author}
  {\bibfnamefont {G.}~\bibnamefont {Xie}}, \bibinfo {author} {\bibfnamefont
  {T.~A.}\ \bibnamefont {Nguyen}}, \bibinfo {author} {\bibfnamefont
  {E.}~\bibnamefont {Karimi}}, \bibinfo {author} {\bibfnamefont
  {L.}~\bibnamefont {Marrucci}}, \bibinfo {author} {\bibfnamefont {D.~A.}\
  \bibnamefont {Nolan}}, \bibinfo {author} {\bibfnamefont {R.~R.}\ \bibnamefont
  {Alfano}},\ and\ \bibinfo {author} {\bibfnamefont {A.~E.}\ \bibnamefont
  {Willner}},\ }\bibfield  {title} {\bibinfo {title} {Gbit/s mode division
  multiplexing over free space using vector modes and a q-plate mode
  (de)multiplexer},\ }\href {https://doi.org/10.1364/OL.40.001980} {\bibfield
  {journal} {\bibinfo  {journal} {Opt. Lett.}\ }\textbf {\bibinfo {volume}
  {40}},\ \bibinfo {pages} {1980} (\bibinfo {year} {2015})}\BibitemShut
  {NoStop}%
\bibitem [{\citenamefont {Firth}\ and\ \citenamefont
  {Skryabin}(1997)}]{intro35}%
  \BibitemOpen
  \bibfield  {author} {\bibinfo {author} {\bibfnamefont {W.~J.}\ \bibnamefont
  {Firth}}\ and\ \bibinfo {author} {\bibfnamefont {D.~V.}\ \bibnamefont
  {Skryabin}},\ }\bibfield  {title} {\bibinfo {title} {Optical solitons
  carrying orbital angular momentum},\ }\href
  {https://doi.org/10.1103/PhysRevLett.79.2450} {\bibfield  {journal} {\bibinfo
   {journal} {Phys. Rev. Lett.}\ }\textbf {\bibinfo {volume} {79}},\ \bibinfo
  {pages} {2450} (\bibinfo {year} {1997})}\BibitemShut {NoStop}%
\bibitem [{\citenamefont {Desyatnikov}\ and\ \citenamefont
  {Kivshar}(2001)}]{intro36}%
  \BibitemOpen
  \bibfield  {author} {\bibinfo {author} {\bibfnamefont {A.~S.}\ \bibnamefont
  {Desyatnikov}}\ and\ \bibinfo {author} {\bibfnamefont {Y.~S.}\ \bibnamefont
  {Kivshar}},\ }\bibfield  {title} {\bibinfo {title} {Necklace-ring vector
  solitons},\ }\href {https://doi.org/10.1103/PhysRevLett.87.033901} {\bibfield
   {journal} {\bibinfo  {journal} {Phys. Rev. Lett.}\ }\textbf {\bibinfo
  {volume} {87}},\ \bibinfo {pages} {033901} (\bibinfo {year}
  {2001})}\BibitemShut {NoStop}%
\bibitem [{\citenamefont {Zhu}\ \emph {et~al.}(2018)\citenamefont {Zhu},
  \citenamefont {Chen}, \citenamefont {Li}, \citenamefont {Zhao}, \citenamefont
  {Zhou}, \citenamefont {Hu}, \citenamefont {Gao}, \citenamefont {Lu},\ and\
  \citenamefont {Shi}}]{intro37}%
  \BibitemOpen
  \bibfield  {author} {\bibinfo {author} {\bibfnamefont {Z.-H.}\ \bibnamefont
  {Zhu}}, \bibinfo {author} {\bibfnamefont {P.}~\bibnamefont {Chen}}, \bibinfo
  {author} {\bibfnamefont {H.-W.}\ \bibnamefont {Li}}, \bibinfo {author}
  {\bibfnamefont {B.}~\bibnamefont {Zhao}}, \bibinfo {author} {\bibfnamefont
  {Z.-Y.}\ \bibnamefont {Zhou}}, \bibinfo {author} {\bibfnamefont
  {W.}~\bibnamefont {Hu}}, \bibinfo {author} {\bibfnamefont {W.}~\bibnamefont
  {Gao}}, \bibinfo {author} {\bibfnamefont {Y.-Q.}\ \bibnamefont {Lu}},\ and\
  \bibinfo {author} {\bibfnamefont {B.-S.}\ \bibnamefont {Shi}},\ }\bibfield
  {title} {\bibinfo {title} {Fragmentation of twisted light in photon--phonon
  nonlinear propagation},\ }\href {https://doi.org/10.1063/1.5020082}
  {\bibfield  {journal} {\bibinfo  {journal} {Applied Physics Letters}\
  }\textbf {\bibinfo {volume} {112}},\ \bibinfo {pages} {161103} (\bibinfo
  {year} {2018})}\BibitemShut {NoStop}%
\bibitem [{\citenamefont {Bigelow}\ \emph {et~al.}(2004)\citenamefont
  {Bigelow}, \citenamefont {Zerom},\ and\ \citenamefont {Boyd}}]{intro38}%
  \BibitemOpen
  \bibfield  {author} {\bibinfo {author} {\bibfnamefont {M.~S.}\ \bibnamefont
  {Bigelow}}, \bibinfo {author} {\bibfnamefont {P.}~\bibnamefont {Zerom}},\
  and\ \bibinfo {author} {\bibfnamefont {R.~W.}\ \bibnamefont {Boyd}},\
  }\bibfield  {title} {\bibinfo {title} {Breakup of ring beams carrying orbital
  angular momentum in sodium vapor},\ }\href
  {https://doi.org/10.1103/PhysRevLett.92.083902} {\bibfield  {journal}
  {\bibinfo  {journal} {Phys. Rev. Lett.}\ }\textbf {\bibinfo {volume} {92}},\
  \bibinfo {pages} {083902} (\bibinfo {year} {2004})}\BibitemShut {NoStop}%
\bibitem [{\citenamefont {Izdebskaya}\ \emph {et~al.}(2012)\citenamefont
  {Izdebskaya}, \citenamefont {Rebling}, \citenamefont {Desyatnikov},\ and\
  \citenamefont {Kivshar}}]{intro39}%
  \BibitemOpen
  \bibfield  {author} {\bibinfo {author} {\bibfnamefont {Y.~V.}\ \bibnamefont
  {Izdebskaya}}, \bibinfo {author} {\bibfnamefont {J.}~\bibnamefont {Rebling}},
  \bibinfo {author} {\bibfnamefont {A.~S.}\ \bibnamefont {Desyatnikov}},\ and\
  \bibinfo {author} {\bibfnamefont {Y.~S.}\ \bibnamefont {Kivshar}},\
  }\bibfield  {title} {\bibinfo {title} {Observation of vector solitons with
  hidden vorticity},\ }\href {https://doi.org/10.1364/OL.37.000767} {\bibfield
  {journal} {\bibinfo  {journal} {Opt. Lett.}\ }\textbf {\bibinfo {volume}
  {37}},\ \bibinfo {pages} {767} (\bibinfo {year} {2012})}\BibitemShut
  {NoStop}%
\bibitem [{\citenamefont {Qiufang}\ \emph {et~al.}(2020)\citenamefont
  {Qiufang}, \citenamefont {Rongfu}, \citenamefont {Sitong}, \citenamefont
  {Guanxue},\ and\ \citenamefont {Xiumin}}]{intro40}%
  \BibitemOpen
  \bibfield  {author} {\bibinfo {author} {\bibfnamefont {Z.}~\bibnamefont
  {Qiufang}}, \bibinfo {author} {\bibfnamefont {Z.}~\bibnamefont {Rongfu}},
  \bibinfo {author} {\bibfnamefont {D.}~\bibnamefont {Sitong}}, \bibinfo
  {author} {\bibfnamefont {W.}~\bibnamefont {Guanxue}},\ and\ \bibinfo {author}
  {\bibfnamefont {G.}~\bibnamefont {Xiumin}},\ }\bibfield  {title} {\bibinfo
  {title} {Focusing pattern of axisymmetric bessel--gaussian beam with helical
  polarization under triangular modulation},\ }\href
  {https://doi.org/10.1364/AO.376445} {\bibfield  {journal} {\bibinfo
  {journal} {Appl. Opt.}\ }\textbf {\bibinfo {volume} {59}},\ \bibinfo {pages}
  {648} (\bibinfo {year} {2020})}\BibitemShut {NoStop}%
\bibitem [{\citenamefont {Lu}\ \emph {et~al.}(2023)\citenamefont {Lu},
  \citenamefont {Li}, \citenamefont {Zhang}, \citenamefont {Li},\ and\
  \citenamefont {Gao}}]{intro41}%
  \BibitemOpen
  \bibfield  {author} {\bibinfo {author} {\bibfnamefont {C.}~\bibnamefont
  {Lu}}, \bibinfo {author} {\bibfnamefont {J.}~\bibnamefont {Li}}, \bibinfo
  {author} {\bibfnamefont {H.}~\bibnamefont {Zhang}}, \bibinfo {author}
  {\bibfnamefont {S.}~\bibnamefont {Li}},\ and\ \bibinfo {author}
  {\bibfnamefont {X.}~\bibnamefont {Gao}},\ }\bibfield  {title} {\bibinfo
  {title} {Focusing pattern of the laguerre-gaussian beam with polarization
  mixing helical-conical phase modulation},\ }\href
  {https://doi.org/10.1364/JOSAA.492633} {\bibfield  {journal} {\bibinfo
  {journal} {J. Opt. Soc. Am. A}\ }\textbf {\bibinfo {volume} {40}},\ \bibinfo
  {pages} {1303} (\bibinfo {year} {2023})}\BibitemShut {NoStop}%
\bibitem [{\citenamefont {Daloi}\ and\ \citenamefont {Dey}(2022)}]{intro42}%
  \BibitemOpen
  \bibfield  {author} {\bibinfo {author} {\bibfnamefont {N.}~\bibnamefont
  {Daloi}}\ and\ \bibinfo {author} {\bibfnamefont {T.~N.}\ \bibnamefont
  {Dey}},\ }\bibfield  {title} {\bibinfo {title} {Vector beam polarization
  rotation control using resonant magneto optics},\ }\href
  {https://doi.org/10.1364/OE.458390} {\bibfield  {journal} {\bibinfo
  {journal} {Opt. Express}\ }\textbf {\bibinfo {volume} {30}},\ \bibinfo
  {pages} {21894} (\bibinfo {year} {2022})}\BibitemShut {NoStop}%
\bibitem [{\citenamefont {Proite}\ \emph {et~al.}(2008)\citenamefont {Proite},
  \citenamefont {Unks}, \citenamefont {Green},\ and\ \citenamefont
  {Yavuz}}]{intro43}%
  \BibitemOpen
  \bibfield  {author} {\bibinfo {author} {\bibfnamefont {N.~A.}\ \bibnamefont
  {Proite}}, \bibinfo {author} {\bibfnamefont {B.~E.}\ \bibnamefont {Unks}},
  \bibinfo {author} {\bibfnamefont {J.~T.}\ \bibnamefont {Green}},\ and\
  \bibinfo {author} {\bibfnamefont {D.~D.}\ \bibnamefont {Yavuz}},\ }\bibfield
  {title} {\bibinfo {title} {Observation of raman self-focusing in an
  alkali-metal vapor cell},\ }\href
  {https://doi.org/10.1103/PhysRevA.77.023819} {\bibfield  {journal} {\bibinfo
  {journal} {Phys. Rev. A}\ }\textbf {\bibinfo {volume} {77}},\ \bibinfo
  {pages} {023819} (\bibinfo {year} {2008})}\BibitemShut {NoStop}%
\bibitem [{\citenamefont {Zhu}\ \emph {et~al.}(2014)\citenamefont {Zhu},
  \citenamefont {Deng}, \citenamefont {Hagley},\ and\ \citenamefont
  {Huang}}]{intro44}%
  \BibitemOpen
  \bibfield  {author} {\bibinfo {author} {\bibfnamefont {C.~J.}\ \bibnamefont
  {Zhu}}, \bibinfo {author} {\bibfnamefont {L.}~\bibnamefont {Deng}}, \bibinfo
  {author} {\bibfnamefont {E.~W.}\ \bibnamefont {Hagley}},\ and\ \bibinfo
  {author} {\bibfnamefont {G.~X.}\ \bibnamefont {Huang}},\ }\bibfield  {title}
  {\bibinfo {title} {Optical self-focusing effect in coherent light scattering
  with condensates},\ }\href {https://doi.org/10.1088/1054-660X/24/6/065402}
  {\bibfield  {journal} {\bibinfo  {journal} {Laser Physics}\ }\textbf
  {\bibinfo {volume} {24}},\ \bibinfo {pages} {065402} (\bibinfo {year}
  {2014})}\BibitemShut {NoStop}%
\bibitem [{\citenamefont {Wang}\ \emph {et~al.}(2019)\citenamefont {Wang},
  \citenamefont {Ji}, \citenamefont {Deng}, \citenamefont {Li}, \citenamefont
  {Wang}, \citenamefont {Fan},\ and\ \citenamefont {Yu}}]{intro45}%
  \BibitemOpen
  \bibfield  {author} {\bibinfo {author} {\bibfnamefont {L.}~\bibnamefont
  {Wang}}, \bibinfo {author} {\bibfnamefont {X.}~\bibnamefont {Ji}}, \bibinfo
  {author} {\bibfnamefont {Y.}~\bibnamefont {Deng}}, \bibinfo {author}
  {\bibfnamefont {X.}~\bibnamefont {Li}}, \bibinfo {author} {\bibfnamefont
  {T.}~\bibnamefont {Wang}}, \bibinfo {author} {\bibfnamefont {X.}~\bibnamefont
  {Fan}},\ and\ \bibinfo {author} {\bibfnamefont {H.}~\bibnamefont {Yu}},\
  }\bibfield  {title} {\bibinfo {title} {Self-focusing effect on the
  characteristics of airy beams},\ }\href
  {https://doi.org/https://doi.org/10.1016/j.optcom.2019.02.058} {\bibfield
  {journal} {\bibinfo  {journal} {Optics Communications}\ }\textbf {\bibinfo
  {volume} {441}},\ \bibinfo {pages} {190} (\bibinfo {year}
  {2019})}\BibitemShut {NoStop}%
\bibitem [{\citenamefont {Morandotti}\ \emph {et~al.}(2001)\citenamefont
  {Morandotti}, \citenamefont {Eisenberg}, \citenamefont {Silberberg},
  \citenamefont {Sorel},\ and\ \citenamefont {Aitchison}}]{intro46}%
  \BibitemOpen
  \bibfield  {author} {\bibinfo {author} {\bibfnamefont {R.}~\bibnamefont
  {Morandotti}}, \bibinfo {author} {\bibfnamefont {H.~S.}\ \bibnamefont
  {Eisenberg}}, \bibinfo {author} {\bibfnamefont {Y.}~\bibnamefont
  {Silberberg}}, \bibinfo {author} {\bibfnamefont {M.}~\bibnamefont {Sorel}},\
  and\ \bibinfo {author} {\bibfnamefont {J.~S.}\ \bibnamefont {Aitchison}},\
  }\bibfield  {title} {\bibinfo {title} {Self-focusing and defocusing in
  waveguide arrays},\ }\href {https://doi.org/10.1103/PhysRevLett.86.3296}
  {\bibfield  {journal} {\bibinfo  {journal} {Phys. Rev. Lett.}\ }\textbf
  {\bibinfo {volume} {86}},\ \bibinfo {pages} {3296} (\bibinfo {year}
  {2001})}\BibitemShut {NoStop}%
\bibitem [{\citenamefont {Agarwal}\ and\ \citenamefont {Dey}(2009)}]{lr1}%
  \BibitemOpen
  \bibfield  {author} {\bibinfo {author} {\bibfnamefont {G.~S.}\ \bibnamefont
  {Agarwal}}\ and\ \bibinfo {author} {\bibfnamefont {T.~N.}\ \bibnamefont
  {Dey}},\ }\bibfield  {title} {\bibinfo {title} {Non-electromagnetically
  induced transparency mechanisms for slow light},\ }\href
  {https://doi.org/10.1002/lpor.200810041} {\bibfield  {journal} {\bibinfo
  {journal} {Laser {\&} Photonics Reviews}\ }\textbf {\bibinfo {volume} {3}},\
  \bibinfo {pages} {287} (\bibinfo {year} {2009})}\BibitemShut {NoStop}%
\end{thebibliography}%
\end{document}